\documentclass[longauth]{aa}  
\usepackage{graphicx}
\usepackage{txfonts}
\usepackage[colorlinks = true, linkcolor = blue, urlcolor  = blue, citecolor = blue]{hyperref}
\usepackage{euclid}

\def \s8{{$\sigma_8 \, $}}
\def \om{{$\Omega_{\rm m} \, $}}
\def \logas{{$\logten A_{\rm s} \, $}}

\begin{document} 

   \title{ \Euclid: Effect of sample covariance \\
   on the number counts of galaxy clusters
   \thanks{This paper is published on behalf of the Euclid Consortium}}

   \author{A.~Fumagalli$^{1,2,3}$\thanks{\email{alessandra.fumagalli@inaf.it}} \and  A.~Saro$^{1,2,3,4}$ \and  S.~Borgani$^{1,2,3,4}$ \and  T.~Castro$^{1,2,3,4}$ \and  M.~Costanzi$^{1,2,3}$ \and  P.~Monaco$^{1,2,3,4}$ \and  E.~Munari$^{3}$ \and  E.~Sefusatti$^{1,3,4}$ \and  A.~Amara$^{5}$ \and  N.~Auricchio$^{6}$ \and  A.~Balestra$^{7}$ \and  C.~Bodendorf$^{8}$ \and  D.~Bonino$^{9}$ \and  E.~Branchini$^{10,11,12}$ \and  J.~Brinchmann$^{13,14}$ \and  V.~Capobianco$^{9}$ \and  C.~Carbone$^{15}$ \and  M.~Castellano$^{12}$ \and  S.~Cavuoti$^{16,17,18}$ \and  A.~Cimatti$^{19,20}$ \and  R.~Cledassou$^{21,22}$ \and  C.J.~Conselice$^{23}$ \and  L.~Corcione$^{9}$ \and  A.~Costille$^{24}$ \and  M.~Cropper$^{25}$ \and  H.~Degaudenzi$^{26}$ \and  M.~Douspis$^{27}$ \and  F.~Dubath$^{26}$ \and  S.~Dusini$^{28}$ \and  A.~Ealet$^{29}$ \and  P.~Fosalba$^{30,31}$ \and  E.~Franceschi$^{6}$ \and  P.~Franzetti$^{15}$ \and  M.~Fumana$^{15}$ \and  B.~Garilli$^{15}$ \and  C.~Giocoli$^{6,32}$ \and  F.~Grupp$^{8,33}$ \and  L.~Guzzo$^{34,35,36}$ \and  S.V.H.~Haugan$^{37}$ \and  H.~Hoekstra$^{38}$ \and  W.~Holmes$^{39}$ \and  F.~Hormuth$^{40}$ \and  K.~Jahnke$^{41}$ \and  A.~Kiessling$^{39}$ \and  M.~Kilbinger$^{42}$ \and  T.~Kitching$^{25}$ \and  M.~K\"ummel$^{33}$ \and  M.~Kunz$^{43}$ \and  H.~Kurki-Suonio$^{44}$ \and  R.~Laureijs$^{45}$ \and  P.~B.~Lilje$^{37}$ \and  I.~Lloro$^{46}$ \and  E.~Maiorano$^{6}$ \and  O.~Marggraf$^{47}$ \and  K.~Markovic$^{39}$ \and  R.~Massey$^{48}$ \and  M.~Meneghetti$^{6,32,49}$ \and  G.~Meylan$^{50}$ \and  L.~Moscardini$^{6,20,51}$ \and  S.M.~Niemi$^{45}$ \and  C.~Padilla$^{52}$ \and  S.~Paltani$^{26}$ \and  F.~Pasian$^{3}$ \and  K.~Pedersen$^{53}$ \and  V.~Pettorino$^{42}$ \and  S.~Pires$^{42}$ \and  M.~Poncet$^{22}$ \and  L.~Popa$^{54}$ \and  L.~Pozzetti$^{6}$ \and  F.~Raison$^{8}$ \and  J.~Rhodes$^{39}$ \and  M.~Roncarelli$^{6,20}$ \and  E.~Rossetti$^{20}$ \and  R.~Saglia$^{8,33}$ \and  R.~Scaramella$^{12,55}$ \and  P.~Schneider$^{47}$ \and  A.~Secroun$^{56}$ \and  G.~Seidel$^{41}$ \and  S.~Serrano$^{30,31}$ \and  C.~Sirignano$^{28,57}$ \and  G.~Sirri$^{32}$ \and  A.N.~Taylor$^{58}$ \and  I.~Tereno$^{59,60}$ \and  R.~Toledo-Moreo$^{61}$ \and  E.A.~Valentijn$^{62}$ \and  L.~Valenziano$^{6,32}$ \and  Y.~Wang$^{63}$ \and  J.~Weller$^{8,33}$ \and  G.~Zamorani$^{6}$ \and  J.~Zoubian$^{56}$ \and  M.~Brescia$^{18}$ \and  G.~Congedo$^{58}$ \and  L.~Conversi$^{64,65}$ \and  S.~Mei$^{66}$ \and  M.~Moresco$^{6,20}$ \and  T.~Vassallo$^{33}$
   }

    \institute{$^{1}$ IFPU, Institute for Fundamental Physics of the Universe, via Beirut 2, 34151 Trieste, Italy\\
$^{2}$ Dipartimento di Fisica - Sezione di Astronomia, Universit\'a di Trieste, Via Tiepolo 11, I-34131 Trieste, Italy\\
$^{3}$ INAF-Osservatorio Astronomico di Trieste, Via G. B. Tiepolo 11, I-34131 Trieste, Italy\\
$^{4}$ INFN, Sezione di Trieste, Via Valerio 2, I-34127 Trieste TS, Italy\\
$^{5}$ Institute of Cosmology and Gravitation, University of Portsmouth, Portsmouth PO1 3FX, UK\\
$^{6}$ INAF-Osservatorio di Astrofisica e Scienza dello Spazio di Bologna, Via Piero Gobetti 93/3, I-40129 Bologna, Italy\\
$^{7}$ INAF-Osservatorio Astronomico di Padova, Via dell'Osservatorio 5, I-35122 Padova, Italy\\
$^{8}$ Max Planck Institute for Extraterrestrial Physics, Giessenbachstr. 1, D-85748 Garching, Germany\\
$^{9}$ INAF-Osservatorio Astrofisico di Torino, Via Osservatorio 20, I-10025 Pino Torinese (TO), Italy\\
$^{10}$ INFN-Sezione di Roma Tre, Via della Vasca Navale 84, I-00146, Roma, Italy\\
$^{11}$ Department of Mathematics and Physics, Roma Tre University, Via della Vasca Navale 84, I-00146 Rome, Italy\\
$^{12}$ INAF-Osservatorio Astronomico di Roma, Via Frascati 33, I-00078 Monteporzio Catone, Italy\\
$^{13}$ Centro de Astrof\'{\i}sica da Universidade do Porto, Rua das Estrelas, 4150-762 Porto, Portugal\\
$^{14}$ Instituto de Astrof\'isica e Ci\^encias do Espa\c{c}o, Universidade do Porto, CAUP, Rua das Estrelas, PT4150-762 Porto, Portugal\\
$^{15}$ INAF-IASF Milano, Via Alfonso Corti 12, I-20133 Milano, Italy\\
$^{16}$ Department of Physics "E. Pancini", University Federico II, Via Cinthia 6, I-80126, Napoli, Italy\\
$^{17}$ INFN section of Naples, Via Cinthia 6, I-80126, Napoli, Italy\\
$^{18}$ INAF-Osservatorio Astronomico di Capodimonte, Via Moiariello 16, I-80131 Napoli, Italy\\
$^{19}$ INAF-Osservatorio Astrofisico di Arcetri, Largo E. Fermi 5, I-50125, Firenze, Italy\\
$^{20}$ Dipartimento di Fisica e Astronomia, Universit\'a di Bologna, Via Gobetti 93/2, I-40129 Bologna, Italy\\
$^{21}$ Institut national de physique nucl\'eaire et de physique des particules, 3 rue Michel-Ange, 75794 Paris C\'edex 16, France\\
$^{22}$ Centre National d'Etudes Spatiales, Toulouse, France\\
$^{23}$ Jodrell Bank Centre for Astrophysics, School of Physics and Astronomy, University of Manchester, Oxford Road, Manchester M13 9PL, UK\\
$^{24}$ Aix-Marseille Univ, CNRS, CNES, LAM, Marseille, France\\
$^{25}$ Mullard Space Science Laboratory, University College London, Holmbury St Mary, Dorking, Surrey RH5 6NT, UK\\
$^{26}$ Department of Astronomy, University of Geneva, ch. d\'Ecogia 16, CH-1290 Versoix, Switzerland\\
$^{27}$ Universit\'e Paris-Saclay, CNRS, Institut d'astrophysique spatiale, 91405, Orsay, France\\
$^{28}$ INFN-Padova, Via Marzolo 8, I-35131 Padova, Italy\\
$^{29}$ Univ Lyon, Univ Claude Bernard Lyon 1, CNRS/IN2P3, IP2I Lyon, UMR 5822, F-69622, Villeurbanne, France\\
$^{30}$ Institute of Space Sciences (ICE, CSIC), Campus UAB, Carrer de Can Magrans, s/n, 08193 Barcelona, Spain\\
$^{31}$ Institut d’Estudis Espacials de Catalunya (IEEC), Carrer Gran Capit\'a 2-4, 08034 Barcelona, Spain\\
$^{32}$ INFN-Sezione di Bologna, Viale Berti Pichat 6/2, I-40127 Bologna, Italy\\
$^{33}$ Universit\"ats-Sternwarte M\"unchen, Fakult\"at f\"ur Physik, Ludwig-Maximilians-Universit\"at M\"unchen, Scheinerstrasse 1, 81679 M\"unchen, Germany\\
$^{34}$ Dipartimento di Fisica "Aldo Pontremoli", Universit\'a degli Studi di Milano, Via Celoria 16, I-20133 Milano, Italy\\
$^{35}$ INFN-Sezione di Milano, Via Celoria 16, I-20133 Milano, Italy\\
$^{36}$ INAF-Osservatorio Astronomico di Brera, Via Brera 28, I-20122 Milano, Italy\\
$^{37}$ Institute of Theoretical Astrophysics, University of Oslo, P.O. Box 1029 Blindern, N-0315 Oslo, Norway\\
$^{38}$ Leiden Observatory, Leiden University, Niels Bohrweg 2, 2333 CA Leiden, The Netherlands\\
$^{39}$ Jet Propulsion Laboratory, California Institute of Technology, 4800 Oak Grove Drive, Pasadena, CA, 91109, USA\\
$^{40}$ von Hoerner \& Sulger GmbH, Schlo{\ss}Platz 8, D-68723 Schwetzingen, Germany\\
$^{41}$ Max-Planck-Institut f\"ur Astronomie, K\"onigstuhl 17, D-69117 Heidelberg, Germany\\
$^{42}$ AIM, CEA, CNRS, Universit\'{e} Paris-Saclay, Universit\'{e} Paris Diderot, Sorbonne Paris Cit\'{e}, F-91191 Gif-sur-Yvette, France\\
$^{43}$ Universit\'e de Gen\`eve, D\'epartement de Physique Th\'eorique and Centre for Astroparticle Physics, 24 quai Ernest-Ansermet, CH-1211 Gen\`eve 4, Switzerland\\
$^{44}$ Department of Physics and Helsinki Institute of Physics, Gustaf H\"allstr\"omin katu 2, 00014 University of Helsinki, Finland\\
$^{45}$ European Space Agency/ESTEC, Keplerlaan 1, 2201 AZ Noordwijk, The Netherlands\\
$^{46}$ NOVA optical infrared instrumentation group at ASTRON, Oude Hoogeveensedijk 4, 7991PD, Dwingeloo, The Netherlands\\
$^{47}$ Argelander-Institut f\"ur Astronomie, Universit\"at Bonn, Auf dem H\"ugel 71, 53121 Bonn, Germany\\
$^{48}$ Institute for Computational Cosmology, Department of Physics, Durham University, South Road, Durham, DH1 3LE, UK\\
$^{49}$ California institute of Technology, 1200 E California Blvd, Pasadena, CA 91125, USA\\
$^{50}$ Observatoire de Sauverny, Ecole Polytechnique F\'ed\'erale de Lau- sanne, CH-1290 Versoix, Switzerland\\
$^{51}$ INFN-Bologna, Via Irnerio 46, I-40126 Bologna, Italy\\
$^{52}$ Institut de F\'{i}sica d’Altes Energies (IFAE), The Barcelona Institute of Science and Technology, Campus UAB, 08193 Bellaterra (Barcelona), Spain\\
$^{53}$ Department of Physics and Astronomy, University of Aarhus, Ny Munkegade 120, DK–8000 Aarhus C, Denmark\\
$^{54}$ Institute of Space Science, Bucharest, Ro-077125, Romania\\
$^{55}$ I.N.F.N.-Sezione di Roma Piazzale Aldo Moro, 2 - c/o Dipartimento di Fisica, Edificio G. Marconi, I-00185 Roma, Italy\\
$^{56}$ Aix-Marseille Univ, CNRS/IN2P3, CPPM, Marseille, France\\
$^{57}$ Dipartimento di Fisica e Astronomia “G.Galilei", Universit\'a di Padova, Via Marzolo 8, I-35131 Padova, Italy\\
$^{58}$ Institute for Astronomy, University of Edinburgh, Royal Observatory, Blackford Hill, Edinburgh EH9 3HJ, UK\\
$^{59}$ Instituto de Astrof\'isica e Ci\^encias do Espa\c{c}o, Faculdade de Ci\^encias, Universidade de Lisboa, Tapada da Ajuda, PT-1349-018 Lisboa, Portugal\\
$^{60}$ Departamento de F\'isica, Faculdade de Ci\^encias, Universidade de Lisboa, Edif\'icio C8, Campo Grande, PT1749-016 Lisboa, Portugal\\
$^{61}$ Universidad Polit\'ecnica de Cartagena, Departamento de Electr\'onica y Tecnolog\'ia de Computadoras, 30202 Cartagena, Spain\\
$^{62}$ Kapteyn Astronomical Institute, University of Groningen, PO Box 800, 9700 AV Groningen, The Netherlands\\
$^{63}$ Infrared Processing and Analysis Center, California Institute of Technology, Pasadena, CA 91125, USA\\
$^{64}$ European Space Agency/ESRIN, Largo Galileo Galilei 1, 00044 Frascati, Roma, Italy\\
$^{65}$ ESAC/ESA, Camino Bajo del Castillo, s/n., Urb. Villafranca del Castillo, 28692 Villanueva de la Ca\~nada, Madrid, Spain\\
$^{66}$ APC, AstroParticule et Cosmologie, Universit\'e Paris Diderot, CNRS/IN2P3, CEA/lrfu, Observatoire de Paris, Sorbonne Paris Cit\'e, 10 rue Alice Domon et L\'eonie Duquet, 75205, Paris Cedex 13, France\\
}

   \date{Received ???; accepted ???}

 
  \abstract
   {}
   {We investigate the contribution of \textit{shot-noise} and \textit{sample variance} to the uncertainty of cosmological parameter constraints inferred from cluster number counts in the context of the \Euclid survey.}
   {By analysing 1000 \Euclid-like light-cones, produced with the PINOCCHIO approximate method, we validate the analytical model of Hu $\&$ Kravtsov 2003 for the covariance matrix, which takes into account both sources of statistical error. Then, we use such covariance to define the likelihood function that better extracts cosmological information from cluster number counts at the level of precision that will be reached by the future \Euclid photometric catalogs of galaxy clusters. We also study the impact of the cosmology dependence of the covariance matrix on the parameter constraints.}
   {The analytical covariance matrix reproduces the variance measured from simulations within the $10$ per cent level; such difference has no sizeable effect on the error of cosmological parameter constraints at this level of statistics. Also, we find that the Gaussian likelihood with cosmology-dependent covariance is the only model that provides an unbiased inference of cosmological parameters without underestimating the errors.}
   {}


    \titlerunning{\Euclid: Effect of sample covariance on the number counts of galaxy clusters}
    \authorrunning{A. Fumagalli et al.}
   \maketitle
%
\section{Introduction}
Galaxy clusters are the most massive gravitationally bound systems in the Universe ($M\sim10^{14}$ -- $10^{15}\,{\rm M_\odot} $) and they are composed of dark matter for $85$ per cent, hot ionized gas for $12$ per cent and stars for $3$ per cent \citep{pratt19}. These massive structures are formed by the gravitational collapse of initial perturbations of the matter density field, through a hierarchical process of accretion and merging of small objects into increasingly massive systems \citep{kravtsov12}.
Therefore galaxy clusters have several properties that can be used to obtain cosmological information on the geometry and the evolution of the large-scale structure of the Universe (LSS). In particular, the abundance and spatial distribution of such objects are sensitive to the variation of several cosmological parameters, such as the RMS mass fluctuation of the (linear) power spectrum on $8\,h^{-1}\,{\rm  Mpc}$ scales (\s8) and the matter content of the Universe (\om) \citep{ borgani99, schuecker03, allen11,pratt19}. Moreover, clusters can be observed at low redshift (out to redshift $z \sim 2$), thus sampling the cosmic epochs during which the effect of dark energy begins to dominate the expansion of the Universe; as such, the evolution of the statistical properties of galaxy clusters should allow us to place constraints on the dark energy equation of state, and then detect possible deviations of dark energy from a simple cosmological constant \citep{sartoris12}. Finally, such observables can be used to constrain neutrino masses \citep[e.g.][]{costanzi13,mantz15, costanzi19, bocquet19,desy1cl}, the Gaussianity of initial conditions \citep[e.g.][]{sartoris10, mana13} and the behavior of gravity on cosmological scales \citep[e.g.][]{cataneo18, bocquet15}.

The main obstacle in the use of clusters as cosmological probes lies in the proper calibration of systematic uncertainties that characterize the analyses of cluster surveys. First, cluster masses are not directly observed but must be inferred through other measurable properties of clusters, e.g. properties of their galaxy population (i.e. richness, velocity dispersion) or of the intracluster gas (i.g., total gas mass, temperature, pressure). The relationships between these observables and clusters masses, called \textit{scaling relations}, provide a statistical measurement of masses, but require an accurate calibration in order to correctly relate the mass proxies with the actual cluster mass. Moreover, scaling relations can be affected by intrinsic scatter due to the properties of individual clusters and baryonic physics effects, that complicate the calibration process \citep{kravtsov12, pratt19}. Other measurement errors are related to the estimation of redshifts and the selection function \citep{allen11}. In addition, there may be theoretical systematics linked to the modelling of the statistical errors: \textit{shot-noise}, the uncertainty due to the discrete nature of data, and \textit{sample variance}, the uncertainty due to the finite size of the survey; in the case of a “full-sky” survey, the latter takes the name of \textit{cosmic variance} and describes the fact that we can observe a single random realization of the Universe \citep[e.g.][]{valageas11}. Finally, the analytical models describing the observed distributions, such as the mass function and halo bias, have to be carefully calibrated, to avoid introducing further systematics \citep[e.g.][]{sheth02, tinker08, tinker10, bocquet15, despali16, castro20}. 

The study and the control of these uncertainties are fundamental for future surveys, which will provide large cluster samples that will allow us to constrain cosmological parameters with a level of precision much higher than that obtained so far. One of the main forthcoming surveys is the European Space Agency (ESA) mission \Euclid\footnote{{\url{http://www.euclid-ec.org}}}, planned for 2022, which will map $\sim 15\,000 \ {\rm deg}^2$ of the extragalactic sky up to redshift 2, in order to investigate the nature of dark energy, dark matter and gravity. Galaxy clusters are among the cosmological probes that will be used by \Euclid: the mission is expected to yield a sample of $\sim 10^5$ clusters using photometric and spectroscopic data and through gravitational lensing \citep{laurejis11, euclid2019}. A forecast of the capability of the \Euclid cluster survey has been performed by \citet{sartoris16}, which shows the effect of the photometric selection function on the number of detected objects and the consequent cosmological constraints for different cosmological models. Also, \citet{kohlinger15} show that the weak lensing systematics in the mass calibration are under control for \Euclid, which will be limited by the cluster samples themselves.

The aim of this work is to assess the contribution of shot-noise and sample variance to the statistical error budget expected for the \Euclid photometric survey of galaxy clusters. The expectation is that the level of shot-noise error would decrease due to the large number of detected clusters, making the sample variance not negligible anymore. To quantify the contribution of these effects, an accurate statistical analysis is required, to be performed on a large number of realizations of past-light cones extracted from cosmological simulations describing the distribution of cluster-sized halos. This is made possible using approximate methods for such simulations \citep[e.g.][ for a review]{monaco16}. A class of these methods describes the formation process of dark matter halos, i.e. the dark matter component of galaxy clusters, through Lagrangian Perturbation Theory (LPT), which provides the distribution of large-scale structures in a faster and computationally less expensive way than through “exact” N-body simulations. As a disadvantage, such catalogs are less accurate and have to be calibrated, in order to reproduce N-body results with sufficient precision. 
By using a large set of LPT-based simulations, we test the accuracy of an analytical model for the computation of the covariance matrix and define which is the best likelihood function to optimize the extraction of unbiased cosmological information from cluster number counts. In addition, we also analyze the impact of the cosmological dependence of the covariance matrix on the estimation of cosmological parameters.

This paper is organized as follows: in Sect. \ref{th_sec} we present the quantities involved in the analysis, such as the mass function, likelihood function and covariance matrix, while in Sect. \ref{sim_sec} we describe the simulations used in this work, which are dark matter halo catalogs produced by the PINOCCHIO algorithm \citep{monaco02, munari17}. In Sect. \ref{res_sec} we present the analyses and the results that we obtain by studying the number counts: in Sect. \ref{cov_sec} (and in Appendix \ref{app_sec}) we validate the analytical model for the covariance matrix, by comparing it with the matrix from simulations. In Sect. \ref{dz_sec} we analyze the effect of the mass and redshift binning on the estimation of parameters, while in Sect. \ref{logl_sec} we compare the effect on the parameter posteriors of different likelihood models. In Sect. \ref{conc_sec} we present our conclusions. While this paper is focused on the analysis relevant for a cluster survey similar in sky coverage and depth to the \Euclid one, for completeness we provide in Appendix \ref{survey_sec} results relevant for present and ongoing surveys. 

\section{Theoretical background} \label{th_sec}
In this section we introduce the theoretical framework needed to model the cluster number counts and derive cosmological constraints via Bayesian inference.

\subsection{Number counts of galaxy clusters} \label{mf_sec}
The starting point to model the number counts of galaxy clusters is given by the halo mass function $\diff n(M,z)$, defined as the comoving volume number density of collapsed objects at redshift $z$ with masses between $M$ and $M+\diff M$ \citep{press74},
\begin{equation}
\label{mf}
\frac{\diff n(M,z)}{\diff \ln M} = \frac{\bar{\rho}_{\rm m}}{M}\, \nu f(\nu)\, \frac{\diff \ln \nu}{\diff \ln M}\,,
\end{equation}
where $\bar{\rho}_{\rm m}/M$ is the inverse of the Lagrangian volume of a halo of mass $M$, and $\nu = \delta_{\rm c}/ \sigma(R,z)$ is the \textit{peak height}, defined in terms of the variance of the linear density field smoothed on scale $R$,
\begin{equation}
\label{sigma2}
\sigma^2(R,z) = \frac{1}{2 \pi^2} \int \diff k \;k^2\, P(k,z)\, W^2_R(k)\,,   
\end{equation}
where $R$ is the radius enclosing the mass $M=\frac{4\pi}{3} \bar{\rho}_{\rm m} R^3$, $W_R(k)$ is the filtering function and $P(k,z)$ the initial matter power spectrum, linearly extrapolated to redshift $z$. The term $\delta_{\rm c}$ represents the critical linear overdensity for the spherical collapse and contains a weak dependence on cosmology and redshift that can be expressed as \citep{nakamura97}
\begin{equation}
\delta_{\rm c}(z) = \frac{3}{20} (12\pi)^{2/3} [1+0.012299 \logten \Omega_{\rm m}(z)]\,.
\end{equation}
One of the main characteristics of the mass function is that, when expressed in terms of the peak height, its shape is nearly universal, meaning that the multiplicity function $\nu f(\nu)$ can be described in terms of a single variable and with the same parameters for all the redshifts and cosmological models \citep{sheth02}. A number of parametrizations have been derived by fitting the mass distribution from N-body simulations \citep{jenkins01, white02, tinker08, watson13}, in order to describe such universality with the highest possible accuracy. At the present time, a fully universal parametrization has not been found yet, and the main differences between the various results reside in the definition of halos, which can be based on the Friends-of-Friends (FoF) and Spherical Overdensity (SO) algorithms \citep[e.g.][]{white01, kravtsov12} or on the dynamical definition of the Splashback radius \citep{diemer17, diemer20}, and in the overdensity at which halos are identified. The need to improve the accuracy and precision in the mass function parametrization is reflected by the differences found in the cosmological parameter estimation, in particular for future surveys such as \Euclid \citep{salvati20, artis21}.
Another way to predict the abundance of halos is the use of emulators, built by fitting the mass function from simulations as a function of cosmology; such emulators are able to reproduce the mass function within few percent accuracy \citep{heitmann16, mcclintock19, bocquet20}. The description of the cluster mass function is further complicated by the presence of baryons, which have to be taken into account when analyzing the observational data; their effect must therefore be included in the calibration of the model \citep[e.g.][]{cui14,velliscig14,bocquet15,castro20}.

In this work, we fix the mass function assuming that the model has been correctly calibrated. The reference mass function that we assume for our analysis is the one by \citep[][hereafter D16]{despali16} \footnote{In D16 the peak height is defined as $\nu = \delta_{\rm c}^2/ \sigma^2(R,z)$; in such case the factor “2” in Eq.\,\eqref{despali_mf} disappears.},
\begin{equation}
\label{despali_mf}
\nu f(\nu) = 2 A\, \left ( 1 + \frac{1}{\nu'^p} \right )\, \left (\frac{\nu'}{2\pi} \right )^{1/2} {\rm e}^{-\nu'/2} \,,
\end{equation}
with $\nu' = a\nu^2$. The values of the parameters are: $A = 0.3298$, $a = 0.7663$, $p = 0.2579$ (“\textit{All z - \Planck cosmology}” case in D16). Comparisons with numerical simulations show departures from the universality described by this model of the order of $5-8\%$, provided that halo masses are computed within the virial overdensity, as predicted by the spherical collapse model. 

Besides the systematic uncertainty due to the fitting model, the mass function is affected by two sources of statistical error (which do not depend on the observational process): shot-noise and sample variance. Shot-noise is the sampling error that arises from the discrete nature of the data and contributes mainly to the high-mass tail of the mass function, where the number of objects is lower, being proportional to the square root of the number counts. On the other hand, sample variance depends only on the size and the shape of the sampled volume; it arises as a consequence of the existence of super-sample Fourier modes, with wavelength exceeding the survey size, that can not be sampled in the analyses of a finite volume survey. Sample variance introduces correlation between different mass and redshift ranges, unlike the shot-noise that affects only objects in the same bin. For currently available data the main contribution to the error comes from shot-noise, while the sample variance term is usually neglected \citep[e.g.][]{mantz15, bocquet19}. Nevertheless, future surveys will provide catalogs with a larger number of objects, making the sample variance comparable, or even greater, than the shot-noise level \citep{hu2003}. 

\subsection{Definition of likelihood functions} \label{l_sec}
The analysis of the mass function is performed through Bayesian inference, by maximizing a likelihood function. The posterior distribution is explored with a Monte Carlo Markov Chains (MCMC) approach \citep{heavens09}, by using a python wrapper for the nested sampler {\fontfamily{qcr}\selectfont PyMultiNest} \citep{buchner14}.

The likelihood commonly adopted in the literature for number counts analyses is the Poissonian one, which takes into account only the shot-noise term. To add the sample variance contribution, the simplest way is to use a Gaussian likelihood. In this work, we considered the following likelihood functions:
\begin{itemize}
    \item Poissonian:
    \begin{equation}
    \label{poiss_l}
        \mathcal{L}(x\,\vert\,\mu) = \prod_{\alpha=1}^{N_z} \prod_{i=1}^{N_M} \frac{\mu_{i\alpha}^{x_{i\alpha}} {\rm e}^{-\mu_{i\alpha}}}{x_{i\alpha}!} \,,
    \end{equation}
    where $x_{i\alpha}$  and $\mu_{i\alpha}$ are, respectively, the observed and expected number counts in the $i$-th mass bin and $\alpha$-th redshift bin. Here the bins are not correlated, since shot-noise does not produce cross-correlation, and the likelihoods are simply multiplied;
    \item Gaussian with shot-noise only:
    \begin{equation}
    \label{gauss_sn_l}
        \mathcal{L}(x\,\vert\,\mu,\,\sigma) =  \prod_{\alpha=1}^{N_z} \prod_{i=1}^{N_M} \frac{\exp{\left \{-\frac{1}{2} (x_{i\alpha}-\mu_{i\alpha})^2/\sigma_{i\alpha}^2\right \}}}{\sqrt{2\pi \sigma_{i\alpha}^2}} \,,
    \end{equation}
    where $\sigma^2_{i\alpha} = \mu_{i\alpha}$ is the shot-noise variance. This function represents the limit of the Poissonian case for large occupancy numbers;
    \item Gaussian with shot-noise and sample variance:
    \begin{equation}
    \label{gauss_cov_l}
        \mathcal{L}(x\,\vert\,\mu,\,C) = \frac{ \exp {\left \{-\frac{1}{2} (\mathbf{x}-\boldsymbol{\mu})^T C^{-1}  (\mathbf{x}-\boldsymbol{\mu}) \right \}}}{\sqrt{2\pi \det[C]}} \,,
    \end{equation}
         where $\mathbf{x}=\{x_{i\alpha}\}$ and $\boldsymbol{\mu}=\{\mu_{i\alpha}\}$, while $C=\{C_{\alpha\beta ij}\}$ is the covariance matrix which correlates different bins due to the sample variance contribution. This function is also valid in the limit of large numbers, as the previous one.
\end{itemize}

We maximise the average likelihood, defined as
\begin{equation}
    \ln \mathcal{L}^{{\rm tot}} = \frac{1}{N_{\rm S}} \sum_{a=1}^{N_{\rm S}} \ln \mathcal{L}^{(a)} \,,
    \label{mean_logl}
\end{equation}
where $N_{\rm S}=1000$ is the number of light-cones and $\ln \mathcal{L}^{(a)}$ is the likelihood of the $a$-th light-cone evaluated according to the equations described above.  
The posteriors obtained in this way are consistent with those of a single light-cone but, in principle, centered on the input parameter values since the effect of cosmic variance that affects each realization of the matter density field is averaged-out when combining all the 1000 light-cones; this procedure makes it easier to observe possible biases in the parameter posteriors due to the presence of systematics.

To estimate the differences on the parameter constraints between the various likelihood models, we quantify the cosmological gain using the figure of merit (FoM hereafter, \citealt{albrecht06}) in the \om-- \s8 plane, defined as
\begin{equation}
\mathrm{FoM}(\Omega_{\rm m }, \sigma_8) = \frac{1}{\sqrt{\det \left [ \mathrm{Cov}(\Omega_{\rm m }, \sigma_8) \right ] }} \,,  
\end{equation}
where Cov(\om, \s8) is the parameter covariance matrix computed from the sampled points in the parameter space. The FoM is proportional to the inverse of the area enclosed by the ellipse representing the $68$ per cent confidence level and gives a measure of the accuracy of the parameter estimation: the larger the FoM, the more precise is the evaluation of the parameters. However, a larger FoM may not indicate a more efficient method of information extraction, but rather an underestimation of the error in the likelihood analysis.

\subsection{Covariance matrix} \label{cov_sec_th}
The covariance matrix can be estimated from a large set of simulations through the equation
\begin{equation}
\label{cov_sim}
C_{\alpha \beta i j} = \frac{1}{N_{\rm S}} \sum_{m=1}^{N_{\rm S}} (n_{i\alpha}^{(m)} - \bar{n}_{i\alpha}) (n_{j\beta}^{(m)} - \bar{n}_{j\beta}) \,,
\end{equation}
where $m = 1,..,N_{\rm S}$ indicates the simulation, $n_{i, \alpha}^{(m)}$ is the number of objects in the $i$-th mass bin and in the $\alpha$-th redshift bin for the $m$-th catalog, while $\bar{n}_{i, \alpha}$ represents the same number averaged over the set of $N_{\rm S}$ simulations.
Such matrix describes both the shot-noise variance, given simply by the number counts in each bin, and the sample variance contribution, or more properly \textit{sample covariance}:
\begin{align}
\label{sn_sim}
& C^{\rm SN}_{\alpha \beta i j} = \bar{n}_{i \alpha} \,\delta_{\alpha \beta} \;\delta_{i j }\,,\\
& C^{\rm SV}_{\alpha \beta i j} = C_{\alpha \beta i j} - C^{\rm SN}_{\alpha \beta i j}\,,
\label{sc_sim}
\end{align}
Actually the precision matrix $C^{-1}$ (which has to be included in Eq.\,\ref{gauss_cov_l}), obtained by inverting Eq.\,\eqref{cov_sim}, is biased due to the noise generated by the finite number of realizations; the inverse matrix must therefore be corrected by a factor \citep{anderson03, hartlap06, taylor13}
\begin{equation}
\label{c_unb}
C^{-1}_{\rm unbiased} = \frac{N_{\rm S}-N_{\rm D}-2}{N_{\rm S} -1} \,C^{-1} \,,
\end{equation}
where $N_{\rm S}$ is the number of catalogs and $N_{\rm D}$ the dimension of the data vector i.e. the total number of bins. 

Although the use of simulations allows us to calculate the covariance in a simple way, numerical estimates of the covariance matrix have some limitations, mainly due to the presence of statistical noise which can only be reduced by increasing the number of catalogs. In addition, simulations make it possible to compute the matrix only at their input cosmology, preventing a fully cosmology-dependent analysis. To overcome these limitations, one can adopt an analytic prescription for the covariance matrix \citep{hu2003,lacasa18,valageas11}. This involves a simplified treatment of non-linearities, so that the validity of this approach must be demonstrated by comparing it with simulations. To this end we consider the analytical model proposed by \citet{hu2003} and validate its predictions against simulated data (see Sect. \ref{cov_sec}). As stated before, the total covariance is given by the sum of the shot-noise variance and the sample covariance,
\begin{equation}
C = C^{\rm SN} +C^{\rm SV}\,.
\end{equation}
According to the model, such terms can be computed as
\begin{align}
& C^{{\rm SN}}_{\alpha \beta i j} = \left \langle N \right \rangle_{\alpha i}  \,\delta_{\alpha \beta} \;\delta_{i j }\,,\\
& C^{{\rm SV}}_{\alpha \beta i j} = \left \langle N b \right \rangle_{\alpha i}  \,\left \langle N b \right \rangle_{\beta j} \,S_{\alpha \beta} \,,
\label{samplecov}
\end{align}
where $\langle N \rangle_{\alpha i}$ and $\langle Nb \rangle_{\alpha i}$ are respectively the expectation values of number counts and  number counts times the halo bias in the $i$-th mass bin and $\alpha$-th redshift bin,
\begin{align}
& \langle N \rangle_{\alpha i} = \Omega_{\rm sky} \int_{\Delta z_{\alpha}} \diff z \;\frac{\diff V}{\diff z\, \diff \Omega} \int_{\Delta M_i} \diff M \;\frac{\diff n}{\diff M}(M,z)\,,\\
& \langle Nb \rangle_{\alpha i} = \Omega_{\rm sky} \int_{\Delta z_{\alpha}} \diff z \;\frac{\diff V}{\diff z\, \diff \Omega} \int_{\Delta M_i} \diff M \;\frac{\diff n}{\diff M}(M,z)\,b(M,z) \,,
\end{align}
with $\Omega_{\rm sky} = 2\pi (1-\cos\theta)$, where $\theta$ is the field-of-view angle of the light-cone, and $b(M,z)$ represents the halo bias as a function of mass and redshift. In the following, we adopt for the halo bias the expression provided by \citet{tinker10}. 
The term $S_{\alpha \beta}$ is the covariance of the linear density field between two redshift bins,
\begin{equation}
S_{\alpha \beta} = D(z_{\alpha}) \,D(z_{\beta}) \int \frac{\diff^3 k}{(2\pi)^3} \;P(k) \,W_{\alpha}(\mathbf{k})  \,W_{\beta}(\mathbf{k}) \,,
\end{equation}
where $D(z)$ is the linear growth rate, $P(k)$ is the linear matter power spectrum at the present time, and $W_{\alpha}(\mathbf{k})$ is the window function of the redshift bin, which depends on the shape of the volume probed. The simplest case is the spherical top-hat window function (see Appendix \ref{app_sec}), while the window function for a redshift slice of a light-cone is given in \citet{costanzi19} and takes the form
\begin{equation}
W_{\alpha}(\mathbf{k}) = \frac{4\pi}{V_{\alpha}} \int_{\Delta z_{\alpha}} \diff z \;\frac{\diff V}{\diff z} \sum_{\ell=0}^{\infty} \sum_{m = -\ell}^{\ell} ({\rm i})^\ell \,j_\ell[k\,r(z)] \,Y_{\ell m}(\hat{\mathbf{k}}) \,K_\ell \,,
\end{equation}
where $\diff V/\diff z$ and $V_{\alpha}$ are respectively the volume per unit redshift and the volume of the slice, which depend on cosmology. Also, in the above equation $j_\ell[k\,r(z)]$ are the spherical Bessel functions, $Y_{\ell m}(\hat{\mathbf{k}})$ are the spherical harmonics, $\hat{\mathbf{k}}$ is the angular part of the wave-vector, and $K_\ell$ are the coefficients of the harmonic expansion, such that
\begin{align*}
&K_\ell = \frac{1}{2\sqrt{\pi}}  \hspace{0.3cm} \text{for} \hspace{0.3cm} \ell=0\,, \\
&K_\ell = \sqrt{\frac{\pi}{2\ell+1}} \frac{P_{\ell-1}(\cos\theta) - P_{\ell+1}(\cos\theta) }{\Omega_{\rm sky}}  \hspace{0.3cm} \text{for} \hspace{0.3cm} \ell \neq 0\,,
\end{align*}
where $P_\ell(\cos\theta)$ are the Legendre polynomials.

\section{Simulations} \label{sim_sec}
The accurate estimation of the statistical uncertainty associated with number counts must be carried out with a large set of simulated catalogs, representing 
different realizations of the Universe. Such large number of synthetic catalogs can hardly be provided by N-body simulations, which are able to produce accurate results  but have high computational costs. Instead, the use of approximate methods, based on perturbative theories, makes it possible to generate a large number of catalogs in a faster and far less computationally expensive way compared to N-body simulations. This comes at the expense of less accurate results: perturbative theories give an approximate description of particle and halo displacements which are computed directly from the initial configuration of the gravitational potential, rather than computing the gravitational interactions at each time step of the simulation \citep[e.g.][]{monaco16, sahni95}. 

PINOCCHIO (PINpointing Orbit-Crossing Collapsed HIerarchical Objects) \citep{monaco02, munari17} is an algorithm which generates dark matter halo catalogs through  Lagrangian Perturbation Theory \citep[LPT, e.g. ][]{moutarde91, buchert92, bouchet95} and ellipsoidal collapse \citep[e.g. ][]{bond96, eisenstein95}, up to third order. The code simulates cubic boxes with periodic boundary conditions, starting from a regular grid on which an initial density field is generated in the same way as in N-body simulations. 
A collapse time is computed for each particle using ellipsoidal collapse. The collapsed particles on the grid are then displaced with LPT to form halos, and halos are finally moved to their final positions by applying again LPT. 
The code is also able to build past-light cones (PLC), by replicating the periodic boxes through an “on-the-fly” process that selects only the halos causally connected with an observer at the present time, once the position of the “observer” and the survey sky area are fixed. This method permits us to generate PLC in a continuous way, i.e. avoiding “piling-up” snapshots at a discrete set of redshifts.

The catalogs generated by PINOCCHIO reproduce within $\sim 5-10$ per cent accuracy the two-point statistics on large scales ($k<0.4\,h\,{\rm Mpc}^{-1}$), the linear bias and the mass function of halos derived from full N-body simulations \citep{munari17}.  The accuracy of these statistics can be further increased by re-scaling the PINOCCHIO halo masses in order to match a specific mass function calibrated against N-body simulations. 
\begin{figure}
    \centering
    \includegraphics[scale=0.56]{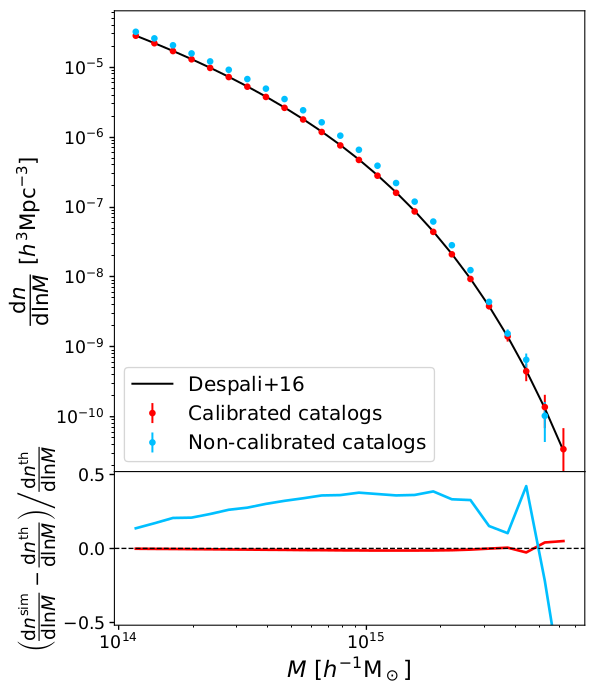}
    \caption{\textit{Top panel}: Comparison between the mass function from the calibrated (red) and the non-calibrated (blue) light-cones, averaged over the 1000 catalogs, in the redshift bin $z=0.1-0.2$;  error bars represent the standard error on the mean. The black line is the D16 mass function. \textit{Bottom panel}: relative difference between the mass function from simulations and the one of D16.}
    \label{dndm_plc}
\end{figure}
\begin{figure}
    \centering
    \includegraphics[scale=0.53]{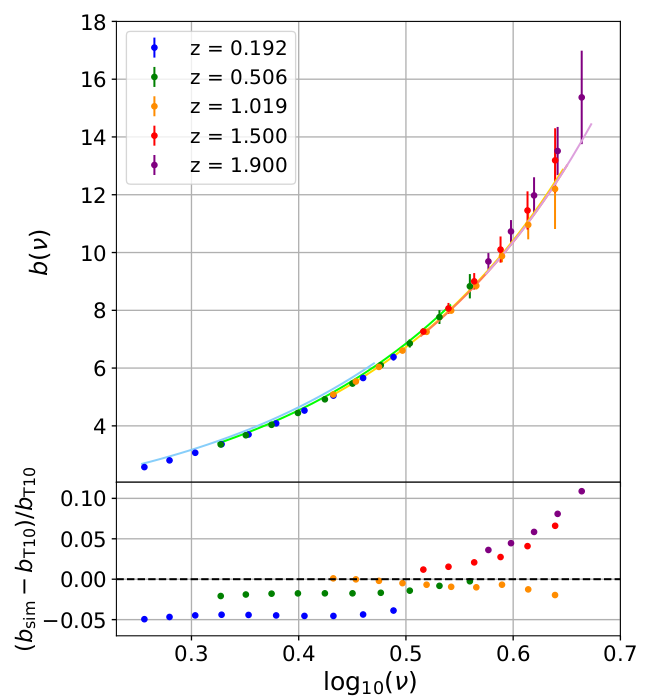}
    \caption{\textit{Top panel}: Halo bias from simulations at different redshifts (colored dots), compared to the analytical model of T10 (lighter solid lines). \textit{Bottom panel}: Fractional differences between the bias from simulations and from the model.}
    \label{t_bias}
\end{figure}

We analyze 1000 past-light-cones\footnote{The PLC can be obtained on request. The list of the available mocks can be found at \url{http://adlibitum.oats.inaf.it/monaco/mocks.html}; the light-cones analyzed are the ones labelled “NewClusterMocks”.}  with aperture of $60^\circ$, i.e. a quarter of the sky, starting from a periodic box of size $L=3870\,h^{-1}\,{\rm  Mpc}$.\footnote{The \Euclid light-cones will be slightly larger than our simulations (about a third of the sky); moreover the survey will cover two separate patches of the sky, which is relevant to the effect of sample variance. However, for this first analysis, the PINOCCHIO light-cones are sufficient to obtain an estimate of the statistical error that will characterize catalogs of such size and number of objects.}
The light-cones cover a redshift range from $z=0$ to $z=2.5$ and contain halos with virial masses above $2.45 \times 10^{13}\,h^{-1}\,{\rm M_\odot} $, sampled with more than 50 particles. The cosmology used in the simulations is the one from \citealt{planck13}: $\Omega_{\rm m } = 0.30711$,  $\Omega_{\rm b} = 0.048254$,  $h = 0.6777$,  $n_{\rm s} = 0.96$,  $\sigma_8 = 0.8288$. 

Before starting to analyze the catalogs, we perform the calibration of halo masses; this step is required both because the PINOCCHIO accuracy in reproducing the halo mass function is “only” 5 percent, and because its calibration has been performed by considering a universal FoF halo mass function, while D16 define halos based on spherical overdensity within the virial radius, demonstrating that the resulting mass function is much nearer to a universal evolution than that of FoF halos.

Masses have been re-scaled by matching the halo mass function of the PINOCCHIO catalogs to the analytical model of D16. More in detail, we predicted the value for each single mass $M_i$ by using the cumulative mass function
\begin{equation}
    N(>M_i) = \Omega_{\rm sky} \int_{\Delta z} \diff z \;\frac{\diff V}{\diff z\, \diff \Omega} \int_{M_i}^\infty \diff M \;\frac{\diff n}{\diff M}(M,z) = i\,,
\end{equation}
where $i=1,2,3..$ and we assigned such values to the simulated halos, previously sorted by preserving the mass order ranking. During this process, all the thousand catalogs were stacked together, which is equivalent to use a 1000 times larger volume: the mean distribution obtained in this way contains fluctuations due to shot-noise and sample variance that are reduced by a factor $\sqrt{1000}$ and can thus be properly compared with the theoretical one, preserving the fluctuations in each rescaled catalog. Otherwise, if the mass function from each single realization was directly compared with the model, the shot-noise and sample variance effects would have been washed away.

In our analyses, we considered objects in the mass range $10^{14}  \le M/{\rm M_\odot} \le 10^{16}$  and redshift range $0 \le z \le 2$; in this interval, each rescaled light-cone contains $
\sim 3 \times 10^5$ halos. We note that this simple constant mass-cut at $10^{14} {\rm M_\odot}$ provides a reasonable approximation to a more refined computation of the mass selection function expected for the \Euclid photometric survey of galaxy clusters (see Fig. 2 of \citealt{sartoris16}; see also \citealt{euclid2019}). 

In Fig.\,\ref{dndm_plc} we show the comparison between the calibrated and non-calibrated mass function of the light-cones, averaged over the 1000 catalogs, in the redshift bin $z = 0.1$ -- 0.2. For a better comparison, in the bottom panel we show the residual between the two mass functions from simulations and the one of D16: while the original distribution clearly differs from the analytical prediction, the calibrated mass function follows the model at all masses, except for some small fluctuations in the high-mass end where the number of objects per bin is low. 

\begin{figure*}
    \centering
    \includegraphics[scale=0.385]{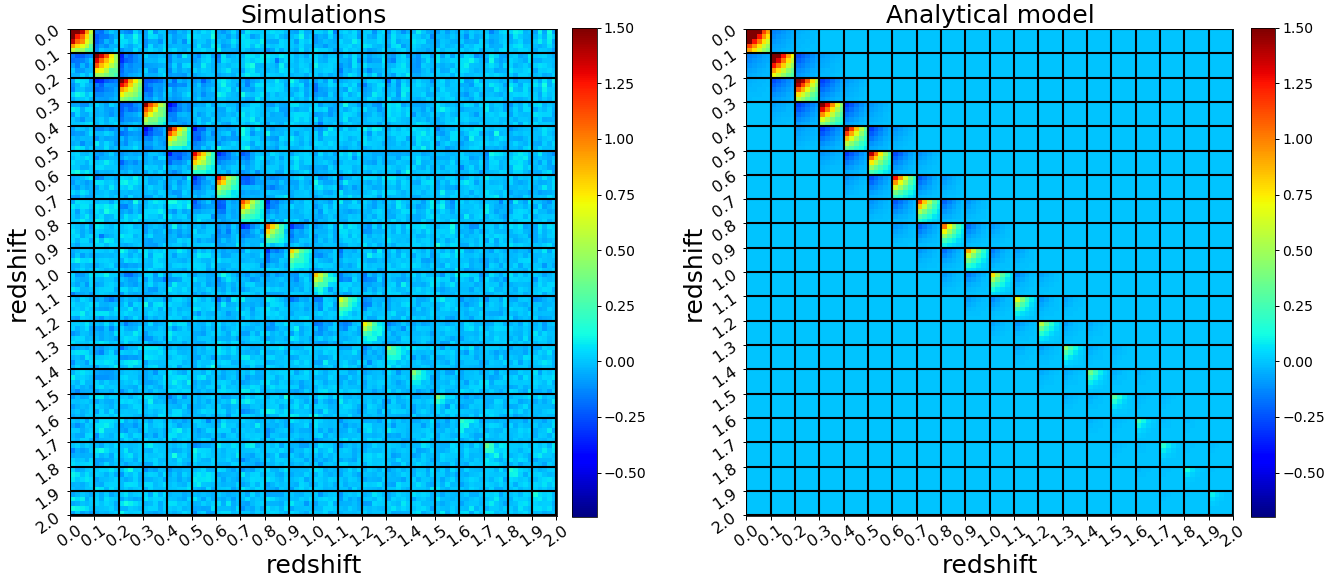}
    \caption{Normalized sample covariance between redshift and mass bins (Eq.\,\ref{norm_cov}), from simulations (\textit{left}) and analytical model (\textit{right}). The matrices are computed in the redshift range $0 \le z \le 1$ with $\Delta z = 0.2$ and the mass range $10^{14} \le M/{\rm M_\odot} \le 10^{16}$, divided in 5 bins. Black lines denote the redshift bins, while in each black square there are different mass bins.}
    \label{cov_plc}
\end{figure*}
We also tested the model for the halo bias of \citet[][hereafter T10]{tinker10}, to understand if the analytical prediction is in agreement with the bias from the rescaled catalogs. The latter is computed by applying the definition 
\begin{equation}
    b^2(\ge M,z) = \frac{\xi_{\rm h}(r,z;M)}{\xi_{\rm m}(r,z)} \,,
\end{equation}
where $\xi_{\rm m}$ is the linear two-point correlation function (2PCF) for matter and $\xi_{\rm h}$ is the 2PCF for halos with masses above a threshold $M$; we use 10 mass thresholds in the range $10^{14} \le M/{\rm M_\odot} \le 10^{15}$. We compute the correlation functions in the range of separations $r = 30$ -- $70 \, h^{-1}\, {\rm Mpc}$, where the approximation of scale-independent bias is valid \citep{manera11}. The error is computed by propagating the uncertainty in $\xi_{\rm h}$, which is an average over the 1000 light-cones. Since the bias from simulations refers to halos with mass $\ge M$, the comparison with the T10 model must be made with an effective bias, i.e. a cumulative bias weighted on the mass function
\begin{equation}
    b_{{\rm eff}} (\ge M,z) = \frac{\int_M^\infty \diff M \;\frac{\diff n}{\diff M}(M,z) \,b(M,z)}{\int_M^\infty \diff M \;\frac{\diff n}{\diff M}(M,z)} \,.
\end{equation}
Such comparison is shown in Fig.\,\ref{t_bias}, representing the effective bias from boxes at various redshifts and the corresponding analytical model, as a function of the peak height (the relation with mass and redshift is shown in Sect.\,\ref{mf_sec}). We notice that the T10 model slightly overestimates (underestimates) the simulated data at low (high) masses and redshifts. However, the difference is below the $5$ per cent level over the whole $\nu$ range, except for high-$\nu$ halos, where the discrepancy is about $10$ per cent, but consistent with our measurements within the error. We conclude that the T10 model can provide a sufficiently accurate description for the halo bias of our simulations.
\begin{figure}
    \centering
    \includegraphics[scale=0.53]{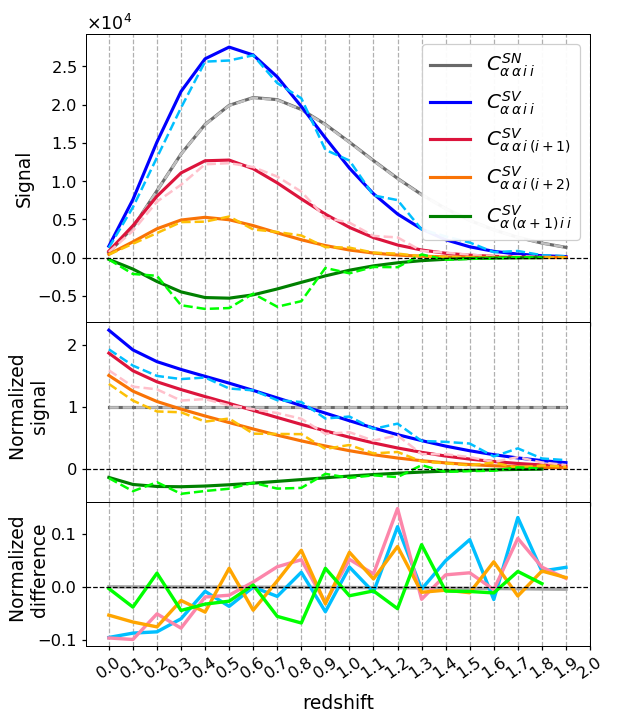}
    \caption{Covariance (\textit{upper panel}) and covariance normalized to the shot-noise level (\textit{central panel}) as predicted by the \citet{hu2003} analytical model (solid lines) and by simulations (dashed lines) for different matrix components: diagonal sample variance terms in blue, diagonal sample variance terms in two different mass bins in red and orange, sample variance between two adjacent redshift bins in green and shot-noise in gray. \textit{Lower panel:} relative difference between analytical model and simulations. The curves are represented as a function of redshift, in the first mass bin ($i = 1$).}
    \label{var_comp}
\end{figure}
\section{Results} \label{res_sec}
In this section we present the results of the covariance comparison and likelihood analyses. First, we validate the analytical covariance matrix, described in Sect. \ref{cov_sec_th}, comparing it with the matrix from the mocks; this allows us to determine whether the analytical model correctly reproduces the results of the simulations.
Once we verified to have a correct description of the covariance, we move to the likelihood analysis. First, we analyse the optimal redshift and mass binning scheme, which will ensure to extract the cosmological information in the best possible way. Then, after fixing the mass and redshift binning scheme, we test the effects on parameter posteriors of different model assumptions: likelihood model, inclusion of sample variance and cosmology dependence. 

With the likelihood analysis, we aim to correctly recover the input values of the cosmological parameters \om, \s8 and \logas. We constrain directly \om and \logas, assuming flat priors in $0.2\,\le\,\Omega_{\rm m}\,\le\,0.4$ and $-9.0\,\le\,\logten A_s\,\le\,-8.0$, and then derive the corresponding value of $\sigma_8$; thus, \s8 and \logas are redundant parameters, linked by the relation $P(k) \propto A_s\,k^{n_s}$ and by Eq.\,\eqref{sigma2}. All the other parameters are set to the \Planck 2014 values. We are interested in detecting possible effects on the results which can occur, in principle, both in terms of biased parameters and over/underestimated parameters errors. The former case indicates the presence of systematics due to an incorrect analysis, while the latter means that not all the relevant sources of error are taken into account.

\subsection{Covariance matrix estimation} \label{cov_sec}
As we mentioned before, the sample variance contribution to the noise can be included in the estimation of cosmological parameters by computing a covariance matrix which takes into account the cross-correlation between objects in different mass or redshift bins. We compute the matrix in the range $0 \le z \le 2$ with $\Delta z = 0.1$ and $10^{14} \le M/{\rm M_\odot} \le 10^{16}$. According to Eq.\,\eqref{c_unb}, since we used $N_S = 1000$ and $N_{\rm D} = 100$ (20 redshift bins and 5 log-equispaced mass bins), we correct the precision matrix by a factor of $0.90$.

In the left panel of Fig.\,\ref{cov_plc} we show the normalized sample covariance matrix, obtained from simulation, which is defined as the relative contribution of the sample variance with respect to the shot-noise level,
\begin{equation}
R^{\rm SV}_{\alpha \beta i j} = \frac{C^{\rm SV}_{\alpha \beta i j}}{\sqrt{\; C^{\rm SN}_{\alpha \alpha i i}\; C^{\rm SN}_{\beta \beta j j}\;}} \,,
\label{norm_cov}
\end{equation}
where $C^{\rm SN}$ and $C^{\rm SV}$ are computed from Eqs.\,\eqref{sn_sim} and \eqref{sc_sim}.
The correlation induced by the sample variance is clearly detected in the block-diagonal covariance matrix (i.e. between mass bins), at least in the low-redshift range where the sample variance contribution is comparable to, or even greater than the shot-noise level. Instead, the off-diagonal and the high-redshift diagonal terms appear affected by the statistical noise mentioned in Sect. \ref{cov_sec_th}, which completely dominates over the weak sample variance (anti-)correlation. 

In the right panel of Fig.\,\ref{cov_plc} we show the same matrix computed with the analytical model: by comparing the two results, we note that the covariance matrix derived from simulations is well reproduced by the analytical model, at least for the diagonal and the first off-diagonal terms, where the former is not dominated by the statistical noise. 
To ease the comparison between simulations and model and between the amount of correlation of the various components, in Fig.\,\ref{var_comp} we show the covariance from model and simulations for different terms and components of the matrix, as a function of redshift: in blue we show the sample variance diagonal terms (i.e. same mass and redshift bin, $C^{{\rm SV}}_{\alpha \alpha i i }$), in red and orange the diagonal sample variance in two different mass bins ($C^{{\rm SV}}_{\alpha \alpha i j}$ with respectively $j=i+1$ and $j=i+2$), in green the sample variance between two adjacent redshift bins ($C^{{\rm SV}}_{\alpha \beta i i}, \,\beta=\alpha+1$) and in gray the shot-noise variance ($C^{{\rm SN}}_{\alpha \alpha i i}$). 
In the upper panel we show the full covariance, in the central panel the covariance normalized as in Eq.\,\eqref{norm_cov} and in the lower panel the normalized difference between model and simulations. Confirming what was noticed from Fig.\,\ref{cov_plc}, the block-diagonal sample variance terms are the dominant sources of error at low redshift, with a signal that rapidly decreases when considering different mass bins (blue, red and orange lines). Shot-noise dominates at high redshift and in the off-diagonal terms. 
We also observe that the cross-correlation between different redshift bins produces a small anti-correlation, whose relevance however seems negligible; further considerations about this point will be presented in Sect. \ref{logl_sec}.

Regarding the comparison between model and simulations, the figure clearly shows that the analytical model reproduces with good agreement the covariance from simulations, with deviations within the 10 per cent level. 
Part of such differences can be ascribed to the statistical noise, which produces random fluctuations in the simulated covariance matrix. 
We also observe, mainly on the block-diagonal terms, a slight underestimation of the correlation at low redshift and a small overestimation at high redshift, which are consistent with the under/overestimation of the T10 halo bias shown in Fig.\,\ref{t_bias}. 
Additional analyses are presented in Appendix \ref{app_sec}, where we treat the description of the model with a spherical top-hat window function. Nevertheless, this discrepancy on the covariance errors has negligible effects on the parameter constraints, at this level of statistics. This comparison will be further analyzed in Sect. \ref{logl_sec}.

\subsection{Redshift and mass binning} \label{dz_sec}
\begin{figure}
    \centering
    \includegraphics[scale=0.54]{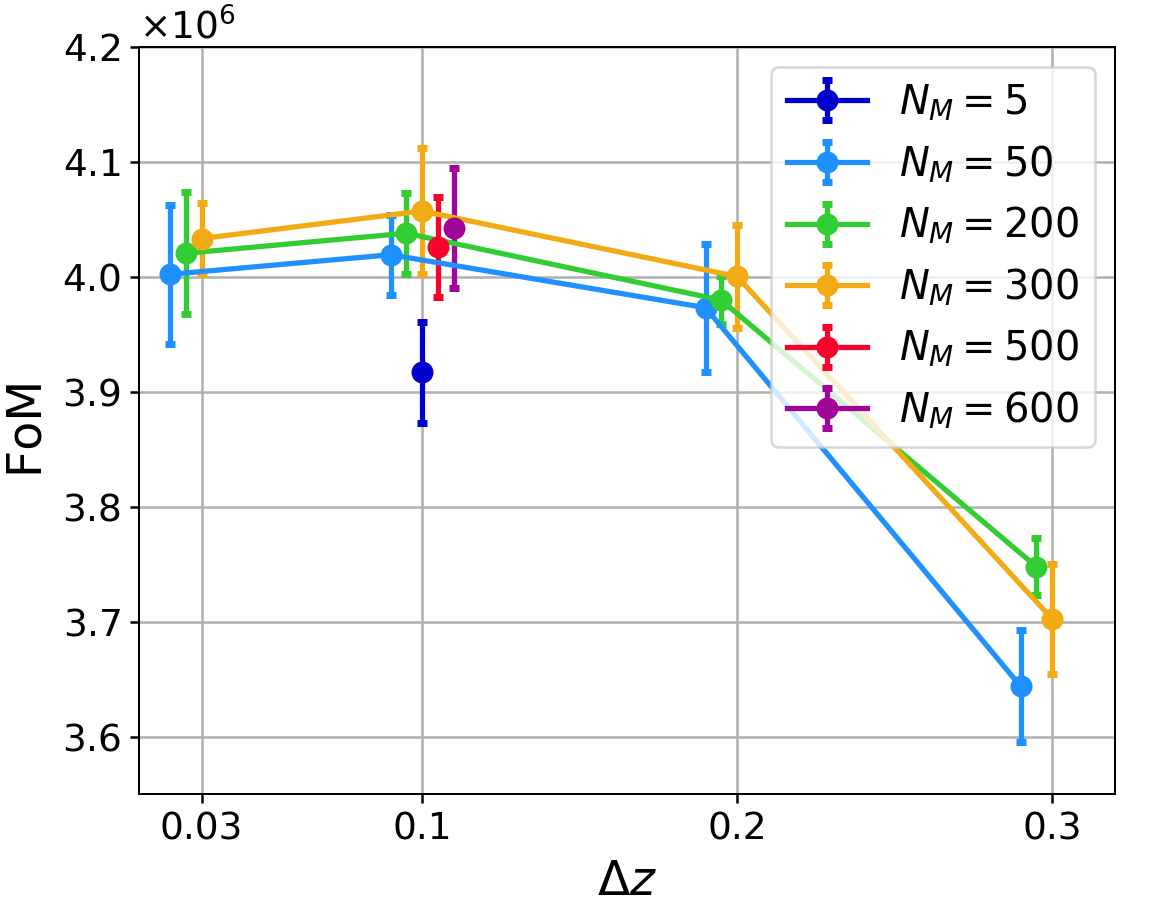} 
    \caption{Figure of merit for the Poissonian likelihood as a function of the redshift bin widths, for different numbers of mass bins. The points represent the average value over 5 realizations and the error bars are the standard error of the mean. A small horizontal offset has been applied to make the comparison clearer.}
    \label{fom_bin_p}
\end{figure}
\begin{figure}
    \centering
    \includegraphics[scale=0.54]{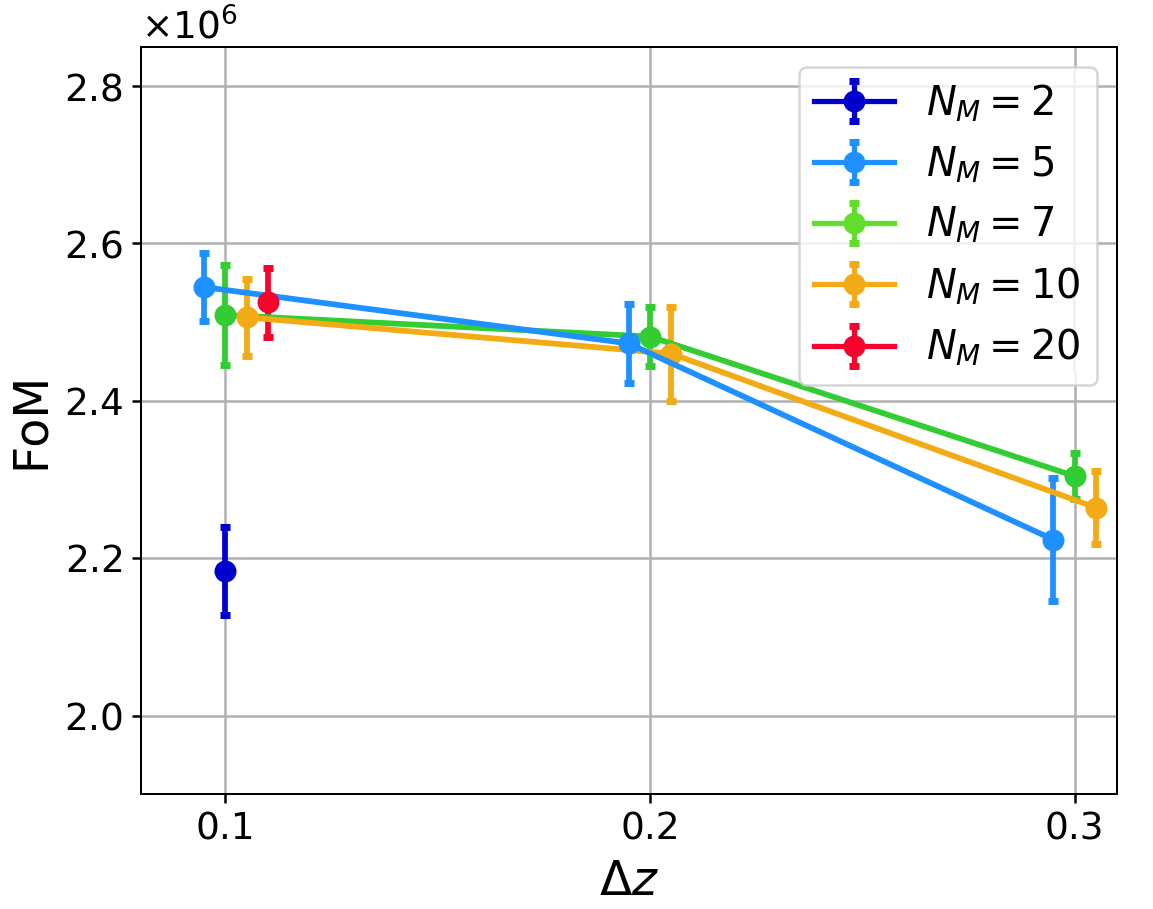} 
    \caption{Same as Fig.\,\ref{fom_bin_p}, for the Gaussian likelihood.}
    \label{fom_bin_g}
\end{figure}
The optimal binning scheme should ensure to extract the maximum information from the data while avoiding wasting computational resources with an exceedingly fine binning: adopting too large bins would hide some information, while too small bins can saturate the extractable information, making the analyses unnecessarily computationally expensive. Moreover, too narrow bins could undermine the validity of the Gaussian approximation due to the low occupancy numbers. This can happen also at high redshift, where the number density of halos drops fast.

To establish the best binning scheme for the Poissonian likelihood function, we analyze the data assuming four redshift bin widths $\Delta z = \{0.03, 0.1, 0.2, 0.3\}$  and three numbers of mass bins $N_M = \{50, 200, 300\}$.
In Fig.\,\ref{fom_bin_p} we show the FoM as a function of $\Delta z$, for different mass binning. Since each result of the likelihood maximization process is affected by some statistical noise, the points represent the mean values obtained from 5 realizations (which are sufficient for a consistent average result), with the corresponding standard error. About the redshift binning, the curve increases with decreasing $\Delta z$ and flattens below $\Delta z \sim 0.2$; from this result we conclude that for bin widths $\lesssim 0.2$ the information is fully preserved and, among these values, we choose $\Delta z = 0.1$ as the bin width that maximize the information. The change of the mass binning affects the results in a minor way, giving points that are consistent with each other for all the redshift bin widths. To better study the effect of the mass binning, we compute the FoM also for $N_M = \{5, 500, 600\}$ at $\Delta z = 0.1$, finding that the amount of recovered information saturates around $N_M = 300$. Thus, we use $N_M = 300$ for the Poissonian likelihood case, corresponding to $\Delta \logten(M/{\rm M_\odot}) = 0.007$.

We repeat the analysis for the Gaussian likelihood (with full covariance), by considering the redshift bin widths $\Delta z = \{0.1, 0.2, 0.3\}$ and three numbers of mass bins $N_M = \{5, 7, 10\}$, plus $N_M = \{2, 20\}$ for $\Delta z = 0.1$. We do not include the case of a tighter redshift or mass binning, to avoid deviating too much from the Gaussian limit of large occupancy numbers. The result for the FoM is shown  Fig.\,\ref{fom_bin_g}, from which we can state that also for the Gaussian case the curve starts to flatten around $\Delta z \sim 0.2$ and $\Delta z\,=\,0.1$ results to be the optimal redshift binning, since for larger bin widths less information is extracted and for tighter bins the number of objects becomes too low for the validity of the Gaussian limit. Also in this case the mass binning does not influence the results in a significant way, provided that the number of binning is not too low. We decide to use $N_M = 5$, corresponding to the mass bin widths $\Delta \logten(M/{\rm M_\odot}) = 0.4$.

\subsection{Likelihood comparison} \label{logl_sec}
\begin{figure}
    \centering
    \includegraphics[scale=0.45]{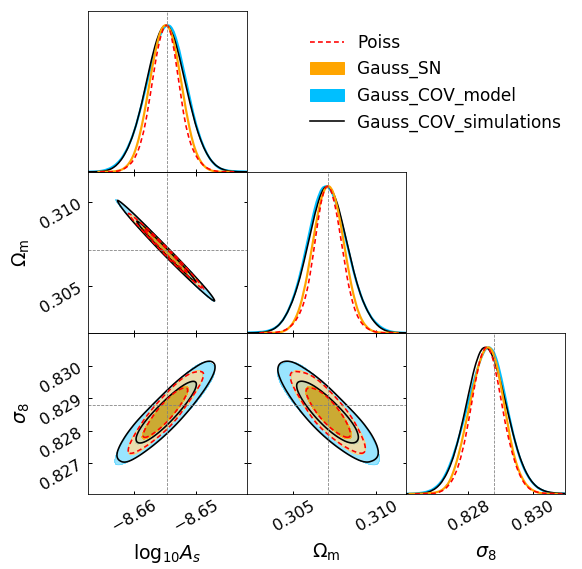}   
    \caption{Contour plots at $68$ and $95$ per cent of confidence level for the three likelihood functions: Poissonian (red), Gaussian with only shot-noise (orange) and Gaussian with shot-noise and sample variance, with covariance from the analytical model (blue) and from simulations (black). The grey dotted lines represent the input values of parameters.}
    \label{post_1}
\end{figure}
\begin{figure}
    \centering
    \includegraphics[scale=0.56]{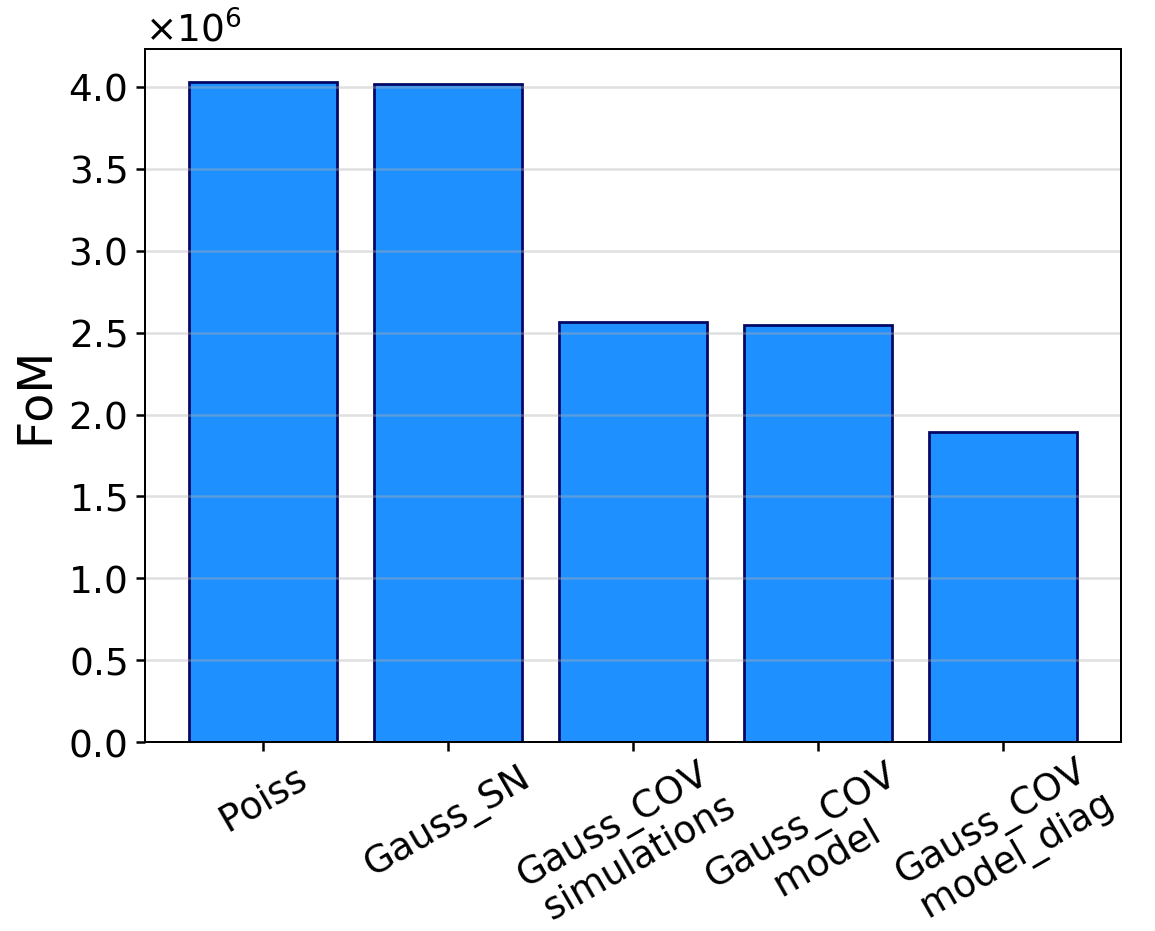}
    \caption{Figure of merit for the different likelihood models: Poissonian, Gaussian with shot-noise, Gaussian with full covariance from simulations, Gaussian with full covariance from the model and Gaussian with block-diagonal covariance from the model. }
    \label{fom_l}
\end{figure}
\begin{figure}
    \centering
    \includegraphics[scale=0.45]{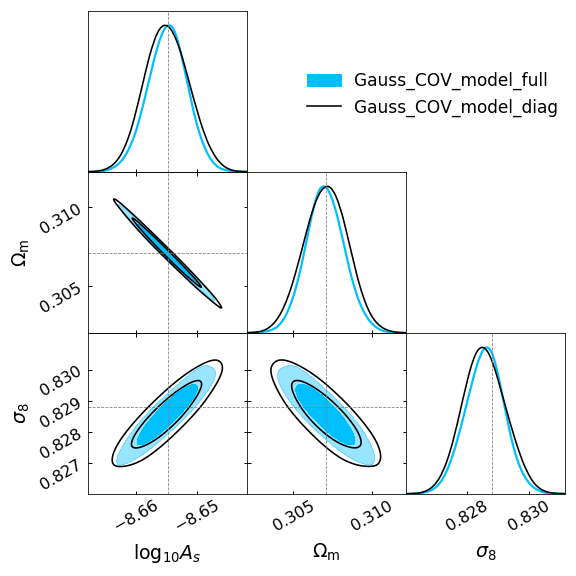}
    \caption{Contour plots at $68$ and $95$ per cent of confidence level for the Gaussian likelihood with full covariance (blue) and the Gaussian likelihood with block-diagonal covariance (black). The grey dotted lines represent the input values of parameters.}
    \label{post_2}
\end{figure}
In this section we present the comparison between the posteriors of cosmological parameters obtained by applying the different definitions of likelihood results on the entire sample of light-cones, by considering the average likelihood defined by Eq.\,\eqref{mean_logl}.

The first result is shown in Fig.\,\ref{post_1}, which represents the posteriors derived from the three likelihood functions: Poissonian, Gaussian with only shot-noise and Gaussian with shot-noise and sample variance (Eqs.\,\ref{poiss_l}, \ref{gauss_sn_l} and \ref{gauss_cov_l}, respectively). For the latter we compute the analytical covariance matrix at the input cosmology and compare it with the results obtained by using the covariance matrix from simulations. The corresponding FoM in the \s8 -- \om plane is shown in 
Fig.\,\ref{fom_l}.
The first two cases look almost the same, meaning that a finer mass binning as the one adopted in the Poisson likelihood does not improve the constraining power compared to the results from a Gaussian plus shot-noise covariance. 
In contrast, the inclusion of the sample covariance (blue and black contours) produces wider contours (and smaller FoM), indicating that neglecting this effect leads to an underestimation of the error on the parameters. Also, there is no significant difference in using the covariance matrix from simulations or the analytical model, since the difference in the FoM is below the percent level. This result means that the level of accuracy reached by the model is sufficient to obtain an unbiased estimation of parameters in a survey of galaxy clusters having sky coverage and cluster statistics comparable to that of the \Euclid survey. According to this conclusion, we will use the analytical covariance matrix to describe the statistical errors for all following likelihood evaluations.

Having established that the inclusion of the sample variance has a non-negligible effect on parameter posteriors, we focus on the Gaussian likelihood case. 
In Fig.\,\ref{post_2} we show the results obtained by using the full covariance matrix and only the block-diagonal of such matrix ($ C_{ij\alpha \alpha}$), i.e. considering shot-noise and sample variance effects between masses at the same redshift but no correlation between different redshift bins. The resulting contours present small differences, as can be seen from the comparison of the FoM in Fig.\,\ref{fom_l}:  the difference in the FoM between the diagonal and full covariance cases is about one third of the effect generated by the inclusion of the full covariance with respect the only shot-noise cases. This means that, at this level of statistics and for this redshift binning, the main contribution to the sample covariance comes from the correlation between mass bins, while the correlation between redshift bins produces a minor effect on the parameter posteriors.
However, the difference between the two FoMs is not necessarily negligible: for three parameters, a $\sim$25\% change in the FoM corresponds to a potential underestimate of the parameter errorbar by $\sim$10\%. The Euclid Consortium is presently requiring for the likelihood estimation that approximations should introduce a bias in parameter errorbars that is smaller than 10\%, so as not to impact the first significant digit of the error. Because the list of potential systematics at the required precision level is long, one should avoid any oversimplification that alone induces such a sizeable effect. The full covariance is thus required to properly describe the sample variance effect at the \Euclid level of accuracy.

\subsection{Cosmology dependence of covariance}
\begin{figure}
    \centering
    \includegraphics[scale=0.45]{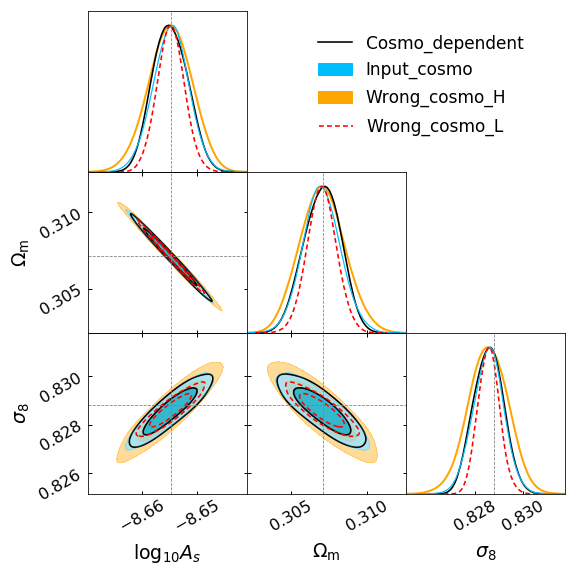} 
    \caption{Contour plots at $68$ and $95$ per of cent confidence level for the Gaussian likelihood evaluated with: a cosmology-dependent covariance matrix (black), a covariance matrix fixed at the input cosmology (blue) and covariance matrices computed at two wrong cosmologies, one with lower parameter values ($\Omega_{\rm m} = 0.295$, $\logten A_{\rm s} = -8.685$ and $\sigma_8 = 0.776$, red) and one with higher parameter values ($\Omega_{\rm m}= 0.320$, $\logten A_{\rm s} = -8.625$ and $\sigma_8 = 0.884$, orange). The grey dotted lines represent the input values of parameters.}
    \label{post_3}
\end{figure}
\begin{figure}
    \centering
    \includegraphics[scale=0.56]{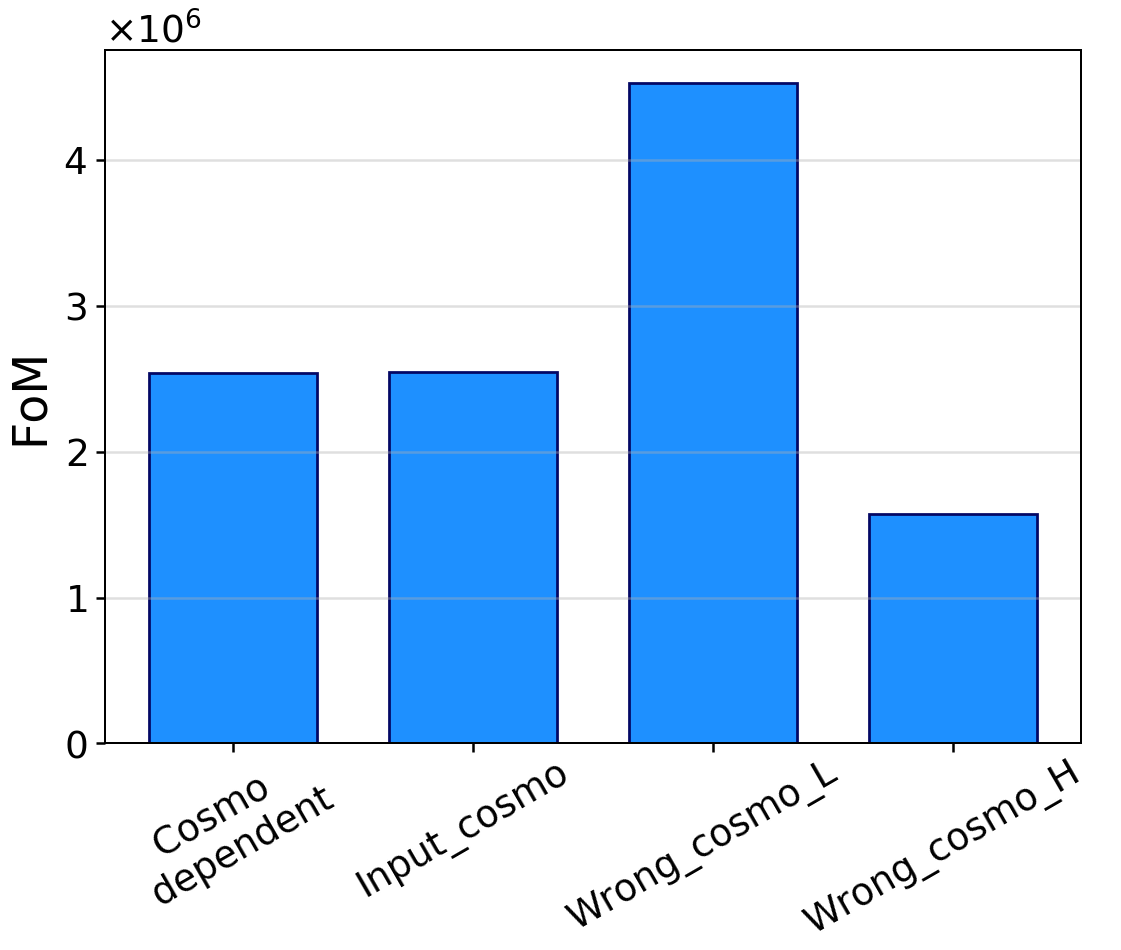}
    \caption{Figure of merit for the models described in Fig.\,\ref{post_3}.}
    \label{fom_cosmo}
\end{figure}
We also investigate if there are differences in using a cosmology-dependent covariance matrix instead of a cosmology-independent one. In fact, the use of a matrix evaluated at a fixed cosmology can represent an advantage, by reducing the computational cost, but may bias the results. In Fig.\,\ref{post_3} we compare the parameters estimated with a cosmology-dependent covariance (black contours), i.e. recomputing the covariance at each step of the MCMC process, with the posteriors obtained by evaluating the matrix at the input cosmology (blue), or assuming a slightly lower/higher value for \om, \logas and \s8  (red and orange contours, respectively), chosen in order to have departures from the fiducial values of the order of $2\sigma$ from \citet{planck18}. Specifically, we fix the parameter values at $\Omega_{\rm m} = 0.295$, $\logten A_{\rm s} = -8.685$ and $\sigma_8 = 0.776$ for the lower case and $\Omega_{\rm m} = 0.320$, $\logten A_{\rm s} = -8.625$ and $\sigma_8 = 0.884$ for the higher case. We notice, also from the FoM comparison in Fig.\,\ref{fom_cosmo}, that there is no appreciable difference between the first two cases. In contrast, when a wrong-cosmology covariance matrix is used we can find either tighter or wider contours, meaning that the effect of shot-noise and sample variance can be either under- or over-estimated. Thus, the use of a cosmology-independent covariance matrix in the analysis of real cluster abundance data might lead to under/overestimated parameter uncertainties at the level of statistic expected for \Euclid.

\section{Discussion and conclusions} \label{conc_sec}
In this work we studied some of the theoretical systematics that can affect the derivation of cosmological constraints from the analysis of number counts of galaxy clusters from a survey having sky-coverage and selection function similar to those expected for the photometric \Euclid cluster survey. One of the aims of the paper was to understand if the inclusion of sample variance, in addition to the shot-noise error, could have some influence on the estimation of cosmological parameters, at the level of statistics that will be reached by the future \Euclid catalogs.
Note that in this work we only consider uncertainties which do not deal with observations, thus neglecting the systematics related to the mass estimation; however \citet{kohlinger15} state that for \Euclid the mass estimates from weak lensing will be under control and, although there will be still additional statistical and systematic uncertainties due to mass calibration, the analysis of real catalogs will approach the ideal case considered here.

To describe the contribution of shot-noise and sample variance, we computed an analytical model for the covariance matrix, representing the correlation between mass and redshift bins as a function of cosmological parameters.
Once the model for the covariance has been properly validated, we moved to the identification of the more appropriate likelihood function to analyse cluster abundance data. The likelihood analysis has been performed with only two free parameters, \om and \logas (and thus $\sigma_8$), since the mass function is less affected by the variation of the other cosmological parameters. 

Both the validation of the analytical model for the covariance matrix and the comparison between posteriors from different likelihood definitions are based on the analysis of an extended set of 1000 \Euclid--like past-light cones generated with the LPT-based PINOCCHIO code \citep{monaco02, munari17}.  

The main results of our analysis can be summarized as follows.

\begin{itemize}
\item To include the sample variance effect in the likelihood analysis, we computed the covariance matrix from a large set of mock catalogs. Most of the sample variance signal is contained in the block-diagonal terms of the matrix, giving a contribution larger than the shot-noise term, at least in the low-mass/low-redshift regime. On the other hand, the anti-correlation between different redshift bins produces a minor effect with respect to the diagonal variance. 

\item  We computed the covariance matrix by applying the analytical model by \citet{hu2003}, assuming the appropriate window function, and verified that it reproduces the matrix from simulations with deviations below the 10 percent accuracy; this difference can be ascribed mainly to the non-perfect match of the T10 halo bias with the one from simulations. However, we verified that such a small difference does not affect the inference of cosmological parameters in a significant way, at the level of statistic of the \Euclid survey. Therefore we conclude that the analytical model of \citet{hu2003} can be reliably applied to compute a cosmology-dependent, noise-free covariance matrix, without requiring a large number of simulations.

\item We established the optimal binning scheme to extract the maximum information from the data, while limiting the computational cost of the likelihood estimation. We analyzed the halo mass function with a Poissonian and a Gaussian likelihood, for different redshift- and mass-bin widths and then computed the figure of merit from the resulting contours in \om-- \s8 plane. The results show that, both for the Poissonian and the Gaussian likelihood, the optimal redshift bin width is $\Delta z = 0.1$: for larger bins, not all the information is extracted, while for smaller bins the Poissonian case saturates and the Gaussian case is no longer a valid approximation. 
The mass binning affects less the results, provided not to choose a too small number of bins. We decided to use $N_M = 300$ for the Poissonian likelihood and $N_M = 5$ for the Gaussian case.

\item We included the covariance matrix in the likelihood analysis and demonstrated that the contribution to the total error budget and the correlation induced by the sample variance term cannot be neglected. In fact, the Poissonian and Gaussian with shot-noise likelihood functions show smaller errorbars with respect to the Gaussian with covariance likelihood, meaning that neglecting the sample covariance leads to an underestimation of the error on parameters, at the \Euclid level of accuracy. As shown in Appendix \ref{survey_sec}, this result holds also for the eROSITA survey, while it is not valid for present surveys like \Planck and SPT. 

\item We verified that the anti-correlation between bins at different redshifts produces a minor, but non-negligible effect on the posteriors of cosmological parameters at the level of statistics reached by the \Euclid survey. We also established that a cosmology-dependent covariance matrix is more appropriate than the cosmology-independent case, which can lead to biased results due to the wrong quantification of shot-noise and sample variance.
\end{itemize}

One of the main results of the analysis presented here is that, for next generation surveys of galaxy clusters, such as \Euclid, sample variance effects need to be properly included, becoming one of the main sources of statistical uncertainty in the cosmological parameters estimation process. The correct description of sample variance is guaranteed by the analytical model validated in this work. 

This analysis represents the first step towards providing all the necessary ingredients for an unbiased estimation of cosmological parameters from the number counts of galaxy clusters. It has to be complemented with the characterization of the other theoretical systematics, e.g. related to the calibration of the halo mass function, and observational systematics, related to the mass-observable relation and to the cluster selection function.

To further improve the extractable information from galaxy clusters, the same analysis will be extended to the clustering of galaxy clusters, by analyzing the covariance of the power spectrum or of the two-point correlation function. Once all the systematics will be calibrated, so as to properly combine such two observables \citep{schuecker03, mana13, lacasa16}, number counts and clustering of galaxy clusters will provide valuable observational constraints, complementary to those of the other two main \Euclid probes, namely galaxy clustering and cosmic shear.

\begin{acknowledgements}
We would like to thank Laura Salvati for useful discussions about the selection functions. SB, AS and AF acknowledge financial support from the ERC-StG 'ClustersxCosmo' grant agreement 716762, the PRIN-MIUR 2015W7KAWC grant, the ASI-Euclid contract and the INDARK grant. TC is supported by the INFN INDARK PD51 grant and by the PRIN-MIUR 2015W7KAWC grant. Our analyses have been carried out at: CINECA, with the projects INA17$\_$C5B32 and IsC82$\_$CosmGC; the computing center of INAF-Osservatorio Astronomico di Trieste, under the coordination of the CHIPP project  \citep{bertocco19, taffoni20}. \AckEC
\end{acknowledgements}

\bibliographystyle{aa} 
\bibliography{biblio}

\begin{appendix}
\section{Covariance on spherical volumes} \label{app_sec} 
We test the \citet{hu2003} model in the simple case of a spherically symmetric survey window function, to quantify the level of agreement between this analytical model and results from LPT-based simulations, before applying it to the more complex geometry of the light-cones. The analytical model is simpler than the one described in Sect. \ref{cov_sec}, as in this case we consider only the correlation between mass bins at the fixed redshift of a PINOCCHIO snapshot; for the sample covariance, Eq.\,\eqref{samplecov} then becomes
\begin{equation}
C^{{\rm SV}}_{ij} = \langle Nb \rangle_i \ \langle Nb \rangle_j \ \sigma^2_R \,,
\end{equation}
where the variance $\sigma^2_R$ is given by Eq.\,\eqref{sigma2}, which contains the Fourier transform of the top-hat window function 
\begin{equation}
\label{tophat}
W_R(k) = 3 \frac{\sin(kR)-kR\cos(kR)}{(kR)^3}.
\end{equation}

The matrix from simulations is obtained by computing spherical random volumes of fixed radius from 1000 periodic boxes of size $L=3870\,h^{-1}\,{\rm  Mpc}$ at a given redshift; the number of spheres was chosen in order to obtain a high number of (statistically) non-overlapping sampling volumes from each box and thus depends on the radius of the spheres. The resulting covariance, computed by applying Eq.\,\eqref{cov_sim} to all sampling spheres, has been compared with the one from the model, with filtering scale \textit{R} equal to the radius of the spheres. 

\begin{figure}
    \centering
    \includegraphics[scale=0.53]{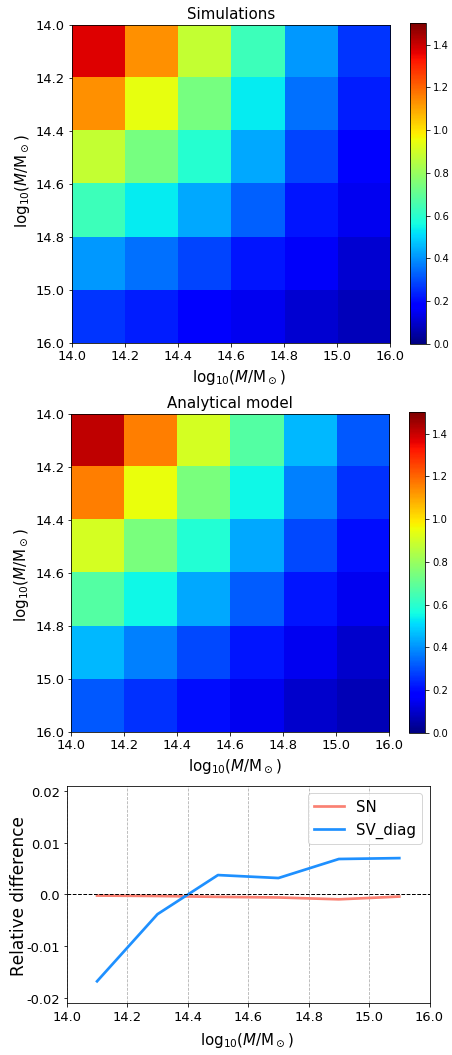}
    \caption{Normalized sample covariance between mass bins from simulations (\textit{top}) and our analytical model (\textit{center}), computed for $10^6$ spherical sub-boxes of radius $R=200 \, h^{-1} {\rm Mpc}$ at redshift $z = 0.506$ and in the mass range $10^{14} \le M/{\rm M_\odot} \le 10^{16}$. In the \textit{bottom panel}, relative difference between simulations and model for the diagonal elements of the sample covariance matrix (blue) and for the shot-noise (red). }
   \label{cov_sb}
\end{figure}

In Fig.\,\ref{cov_sb} we show the resulting normalized matrices computed for $R=200\,h^{-1}\,{\rm  Mpc}$, with $10^3$ sampling spheres for each box. The redshift is $z = 0.506$, and we used 5 log-equispaced mass bins in the range $10^{14} \le M/{\rm M_\odot} \le 10^{15}$ plus one bin for $M = 10^{15}-10^{16}\,{\rm M_\odot}$.  For a better comparison, in the lower panel we show the normalized difference between simulations and model, for the diagonal sample variance terms and for the shot-noise. We notice that the predicted variance is in agreement with the simulated one with a discrepancy less than $2$ per cent. We also notice a slight underestimation of the covariance predicted by the model at low masses and a slight overestimation at high masses. We ascribe this to the modelling of the halo bias, whose accuracy is affected by scatter at the few percent level \citep{tinker10}. 
\begin{figure}
    \centering
    \includegraphics[scale=0.55]{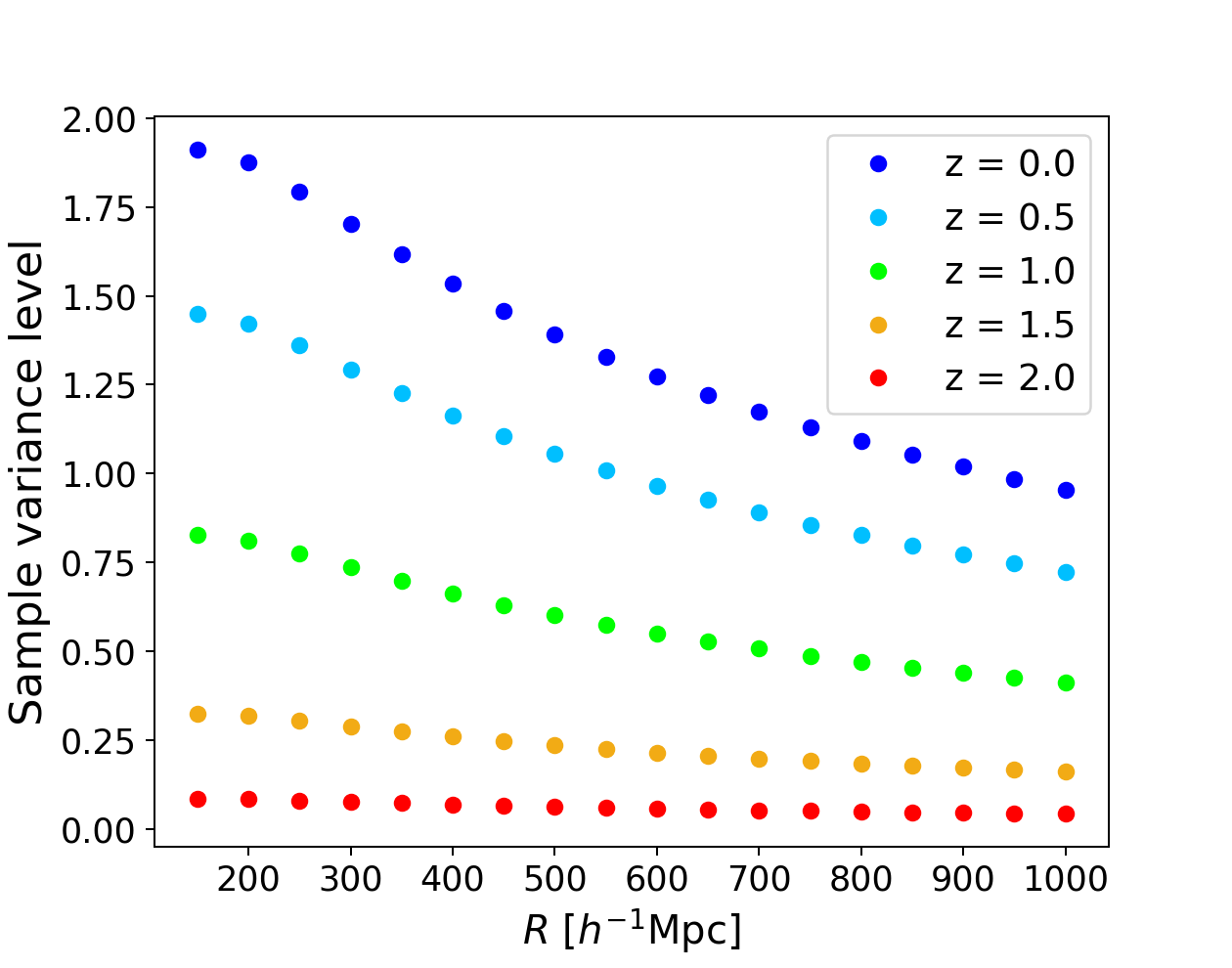}
    \caption{Sample variance level with respect to the shot-noise, in the lowest mass bin, as a function of the filtering scale $R$, at different redshifts.}
   \label{sv_level}
\end{figure}

In Fig.\,\ref{sv_level} we show the (maximum) sample variance contribution relative to the shot-noise level, as a function of the filtering scale, for different redshifts. The curves show that the level of sample variance is lower at high redshift, where the shot-noise dominates due to the small number of objects. Instead, at low redshift ($z < 1$) the sample variance level is even higher than the shot-noise one, and increase as the radius of the spheres decrease; this means that, at least at low redshift where the volumes of the redshift slices in the light-cones are small, such contribution cannot be neglected, not to introduce systematics or underestimate the error on the parameter constraints.

\section{Application to other surveys} \label{survey_sec} 
We repeated the likelihood comparison by mimicking other surveys of galaxy clusters, which differ in their volume sampled and their mass and redshift ranges. More specifically, we consider a \Planck-like \citep{tauber10} and an SPT-like \citep{carlstrom11} cluster survey, both selected through the Sunyaev–Zeldovich effect, which represent two of the main currently available cluster surveys. We also analyse an eROSITA-like \citep{predehl2014} X-ray cluster sample, an upcoming survey that, although not reaching the level of statics that will be provided by \Euclid, will produce a much larger sample than current surveys.

The light-cones have been extracted from our catalogs, by considering the properties (aperture, selection function, redshift range) of the three surveys, as provided by \citet[][see Fig. 4 in their paper]{bocquet16}\footnote{Masses in the paper are defined at the overdensity $\Delta = 500$ with respect to the critical density; the conversion to virial masses has been performed with the python package {\fontfamily{qcr}\selectfont hydro$\_$mc} (\url{https://github.com/aragagnin/hydro_mc}).}. 

The properties of the surveys are as follows:
\begin{description}
    \item[\textbf{SPT-like sample:}] we consider light-cones with an area of $2500 \ {\rm deg}^2$, containing halos with redshifts $z > 0.25$ and masses $M_{500{\rm c}} \ge 3 \times 10^{14}\,{\rm M_\odot}$. We obtain catalogs with $\sim\,1100$ objects. We analyze the redshift range $0.25 \le z \le 1.5$ with bins of width $\Delta z = 0.2$ and the mass range $3\times 10^{14} \le M_{500{\rm c}}/{\rm M_\odot} \le 3\times 10^{15}$, divided in 10 bins for the Poissonian case and in 3 bins for the Gaussian case.\\
    
    \item[\textbf{\Planck-like sample:}] we use the redshift-dependent selection function shown in the reference paper. Since the aperture of the \Planck survey is about twice the size of the \Euclid one, we stack together two light-cones to obtain a \Planck-like light-cone; each of the 500 resulting samples contains $\sim 650$ objects. We consider the redshift range $0 \le z \le 0.8$ with $\Delta z = 0.25$ and mass range  $10^{14} \le M_{\rm vir}/{\rm M_\odot} \le 10^{16}$; the number of mass bins varies for different redshift bins due to the redshift-dependent selection function, and it is chosen in order to have non-empty bins at each redshift (at least 10 objects per bin).\\

    \item[\textbf{eROSITA-like sample:}] we select halos according to the redshift-dependent selection function given by $M_{500{\rm c}}(z)\,\ge\,2.3\,z\,\times\,10^{14}\,{\rm M_\odot}$, with a mass cut at $7 \times 10^{13}\,{\rm M_\odot}$. We analyze the redshift range $0 \le z \le 2$ with $\Delta z=0.1$ and the mass range $10^{14} \le M_{\rm vir}/{\rm M_\odot} \le 10^{16}$ with binning defined in order to have non-empty redshift bins, as for the \Planck case. Also in this case, we stack together four PINOCCHIO light-cones to create a full-sky eROSITA light-cone, obtaining 250 samples containing $\sim 2 \times 10^5$ objects. For the purpose of this analysis we did not include any sensitivity mask, to account for the different depths of different surveyed area, due to the eROSITA scanning strategy.

\end{description}

\begin{figure}
    \centering
    \includegraphics[scale=0.44]{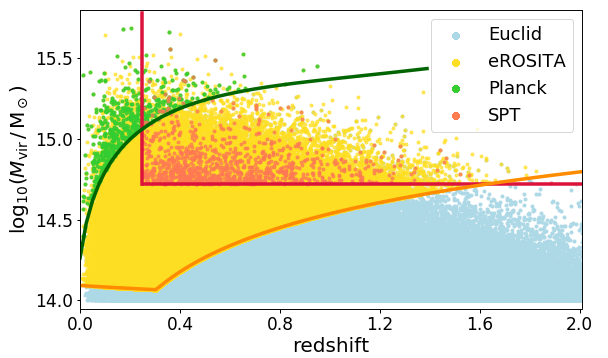}
    \caption{Mass distribution of the three samples extracted from a single light-cone, with the respective selection functions: \Planck in green, SPT in red and eROSITA in orange, overplotted to the full \Euclid sample in blue.} 
    \label{selection}
\end{figure}

In Fig.\,\ref{selection} we show the distribution of cluster masses of the three samples with their selection function, for comparison to the full \Euclid-like catalog. For both SPT and \Planck, despite the different selection functions that favour different mass and redshift ranges, the number of objects is low, so we expect shot-noise to be the main source of uncertainty. In contrast, the eROSITA sample contains a larger number of halos, which should lower the level of shot-noise and make the sample variance non-negligible.

\begin{figure}
    \centering
    \includegraphics[scale=0.45]{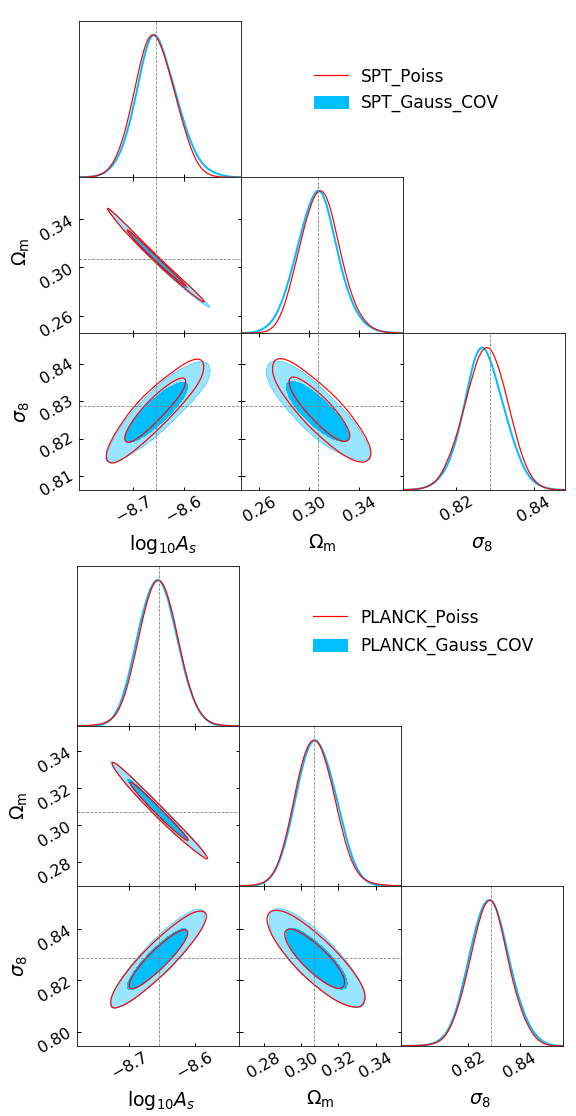}
    \caption{Contour plots at $68$ and $95$ per cent of confidence level for the Poissonian (red) and Gaussian (blue) likelihood for the SPT-like (\textit{top}) and  \Planck-like (\textit{bottom}) samples. The grey dotted lines represent the input values of parameters.}
    \label{post_sp}
\end{figure}

In Fig.\,\ref{post_sp} we show the resulting average contours for the \Planck and SPT samples, obtained with the Poissonian and Gaussian (full covariance) likelihood functions. In both the cases, the contours from the Gaussian case coincide with the Poissonian ones, confirming that for their survey properties, which produce a low number of objects, the shot-noise dominates over the sample variance. Thus, the use of the Poissonian likelihood still represents a good approximation that does not introduce significant differences at the level of statistics given by the present surveys. Moreover, no systematic effects related to uncertainties in the relation between mass and observable (integrated Compton-$y$ parameter in this case), have been included in the analysis. Unlike \Euclid, for these surveys such an uncertainty is expected to dominate the resulting uncertainty on the cosmological parameters \citep{bocquet15}, thus making the choice of the likelihood function conservative, since the posteriors would be larger and the effect of theoretical systematics less significant.

\begin{figure}
    \centering
    \includegraphics[scale=0.45]{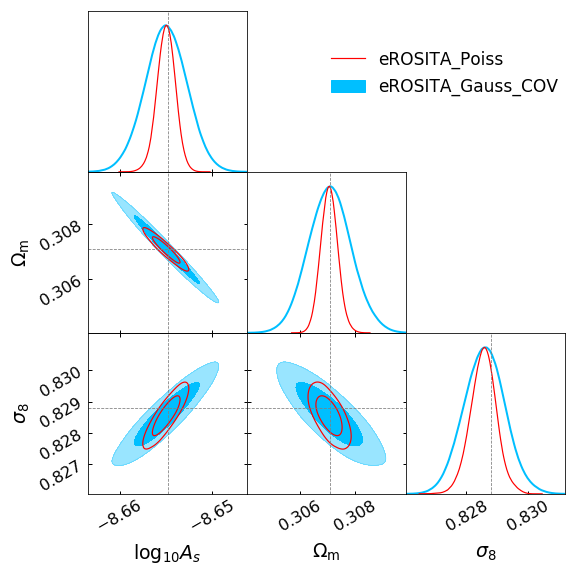}
    \caption{Contour plots at $68$ and $95$ per cent of confidence level for the Poissonian (red) and Gaussian (blue) likelihood for the eROSITA-like sample. The grey dotted lines represent the input values of parameters.}
    \label{eros_post}
\end{figure}

In Fig.\,\ref{eros_post} we show the same result for the eROSITA case. We note that there is a large difference between the Poissonian and the Gaussian case, due to the inclusion of the sample variance effect. Such difference can be ascribed to the mass selection of the survey, which makes the Gaussian contours wider due to the fact that for an X-ray selection, the statistics of counts is dominated by low-redshift/low-mass objects distributed within a relatively small volume, which makes the contribution of sample variance becoming comparable to, or dominant over the shot-noise. 

\end{appendix}
\end{document}